# Making Recommendations Bandwidth Aware

Linqi Song and Christina Fragouli

### Abstract

This paper asks how much we can gain in terms of bandwidth and user satisfaction, if recommender systems became bandwidth aware and took into account not only the user preferences, but also the fact that they may need to serve these users under bandwidth constraints, as is the case over wireless networks. We formulate this as a new problem in the context of index coding: we relax the index coding requirements to capture scenarios where each client has preferences associated with messages. The client is satisfied to receive any message she does not already have, with a satisfaction proportional to her preference for that message. We consistently find, over a number of scenarios we sample, that although the optimization problems are in general NP-hard, significant bandwidth savings are possible even when restricted to polynomial time algorithms.

## I. INTRODUCTION

Recommender systems decide which content to offer to users so as to maximize a benefit (for instance, in advertising networks the benefit could be the profit gained from the ad placement) [2], [3]. These recommendations are currently oblivious to the cost of distributing the content from the server to the points of consumption, which however forms many times the point of failure: unsatisfactory delivery is identified as a core threat to the user experience and has already caused loss of billions of revenue dollars [4]. Wireless consumption in particular, that is increasingly gaining momentum, is inherently subject to bandwidth constraints.

In this paper, we ask: how much could we gain in terms of bandwidth and user satisfaction, if recommendation systems became bandwidth aware, and took into account not only the user preferences, but also the fact that they need to serve these users under bandwidth constraints?

As a first step, we formulate this as a new problem in the context of index coding. The index coding problem [5], [6], [7] considers a server with $m$ messages and $n$ clients. Each client has as side-information a subset of the messages and requires a specific message[1] she does not have. The server can make error-free broadcast transmissions to all clients; the goal is to minimize the number of transmissions so that all clients are satisfied. The error-free broadcasting assumption in index coding offers a first abstraction of modeling wireless, and does not capture all aspects of real world wireless; yet we believe it enables a first understanding of potential trade-offs between benefits and communication cost.

We relax the index coding requirements to capture scenarios where each client has preferences associated with messages: a client can now be satisfied by receiving any message she does not already have; however, the benefit is proportional to how high her preference is for the message she gets. For instance, consider wireless stations serving sale coupons inside a shopping mall: a client walking outside a shop would be happy to receive a coupon she does not already have, but would be happier to receive (and more likely to use) a coupon closer to her interests. We note that the side-information setup fits well with the recommender systems framework [3]: collecting side information about the clients and keeping track of previous content served is an integral part of recommender systems; it is a natural step to leverage this side information, not only to inform recommendations, but to also increase the communication efficiency so as to extract more benefits under communication constraints. But for the amount of interesting work in index coding (eg., [5], [6], [7], [8], [9]), this is the first paper as far as we know that explores trade-offs between user satisfaction and bandwidth.

A challenge we face when setting as our goal to evaluate potential benefits, is that these depend on the preference models we use. There exist numerous models for expressing preferences and for making decisions based on them; clearly we cannot exhaustively investigate all possible preference models. We opt to sample a few models that we think are representative, with the hope of finding consistent trends across them. One model we investigate uses the Borda count method, that has each client rank $m$ messages according to her preferences, and assigns to messages benefits $m, m-1, m-2, \ldots, 1$ with respect to this ranking [10]. We also consider a bimodal preference model, where a fraction of the messages are much more preferable than the remaining. More generally, we considered an arbitrary preference model, where we do not have restrictions on the benefit $w_{ij} \geq 0$ that the message $j$ gives to client $i$.

---

[1] For the case that a user requests multiple messages, it is equivalent to some problem instance where a client requests a single message, as we will discuss in Section II.

To calculate the aggregate benefit, we need to gather the benefits all clients receive from the messages the server transmit. There are also several ways to calculate the benefit for a client when multiple messages are served, such as the sum of the benefits of all received messages, the weighted sum of them, and the maximum among them. In this paper, we focus on the *maximum benefit rule* by counting only the highest-preference message we have served to each client, and discuss other possible rules in Section VII. This is motivated from that, if a client at a certain time can see only one video (or read one article or click one ad), although her device may have downloaded multiple items, she will only see her most preferred one, and we will collect the corresponding benefit. This benefit model aligns well with the index-coding rationale, where only the one message the client wants counts. Moreover, by counting only one best message per client, we limit the potential unfairness of serving multiple messages to the same client and none to others.

The main contribution of this paper is to examine the trade-off between benefit and bandwidth across three scenarios, both theoretically and numerically using designed algorithms. We first provide results for the case where there is no side information for each client and each client has a Borda count preference model over all the messages. We show that the problem is NP-hard, however a simple greedy polynomial time algorithm can achieve an approximation ratio of $1.58$. Moreover, we provide upper and lower bounds of optimal performance as well as an average case analysis, both indicating diminishing returns: the benefit increases with the number of transmissions $t$ only by a multiplicative factor $1 - 1/t$.

The second scenario investigates the case where each client has side-information of the same cardinality and a Borda count preference model over the remaining unknown messages. We prove lower bounds on the optimal benefit, and design a polynomial-time algorithm that achieves a $O(1)$ approximation ratio.

The third scenario is to consider a general case where each client has an arbitrary-sized side information set and we put no restrictions on the preference model except the obvious non-negativity. We show that this problem is hard to approximate within a ratio of $n^{1-\epsilon}$ for any $\epsilon > 0$. In this case, we establish a connection to the Maximum Weighted Independent Set (MWIS) problem, and design a heuristic coded algorithm by leveraging this connection.

We evaluate our algorithms numerically over synthetic and real world data sets (Yahoo!

advertiser bidding data sets [11]). We find that even with one transmission we can in many cases already achieve half of the maximum possible benefit; and in general, we can achieve $80\%$ of the benefit using less than $10\%$ of the transmissions needed to achieve the maximum possible benefit. We also find that leveraging side information, coded transmissions can in some cases enable to double the benefit over uncoded transmissions.

The paper is organized as follows. We formulate the bandwidth aware recommendation problem in Section II, followed by analytically deriving polynomial time algorithms and performance bounds in Sections III-V. We present numerical experiments in Section VI and conclude the paper in Section VII with a discussion of possible extensions of our algorithms.

## II. SETUP AND PROBLEM FORMULATIONS

### A. Setup

We consider that a server has $m$ messages $b_1, b_2, \ldots, b_m$ and $n$ clients $c_1, c_2, \ldots, c_n$. We assume that each message $b_j$, $1 \leq j \leq m$, takes values in a finite field $\mathbb{F}_q$. We will sometimes say message $j$ instead of $b_j$, and similarly, client $i$ instead of $c_i$. We use the notation $[y]$ to denote the set $\{1, 2, \ldots, y\}$ and $|Y|$ to denote the cardinality of the set $Y$ throughout the paper. Each client $i \in [n]$ already has as side information a subset of the $m$ messages; we denote by $S_i \subseteq [m]$ the set of indices of the side information for client $i$ ($S_i$ could be the empty set), and by $R_i = [m] \backslash S_i$ the request set of client $i$, i.e., the indices of messages that the client does not have and may request. The server knows perfectly the side information set of each client and the preference of the clients for each message. We will define the client preferences later in this section.

**Broadcast transmissions and coding:** The server is connected to the clients through an error-free broadcast channel; i.e., all clients perfectly receive each server transmission. We focus on linear schemes in this paper. During the $l$-th transmission, the server transmits $x_l = \sum_{j \in [m]} a_{l,j} b_j$, where $a_{l,j} \in \mathbb{F}_q$ are constant coefficients and the addition/multiplication operations are performed in $\mathbb{F}_q$. Thus, each time, the server transmits either one of the uncoded messages $b_j$, or a linear combination of some of the messages. Assume that the server broadcasts $t$ transmissions $x_1, x_2, \ldots, x_t$, and we denote by a vector $\boldsymbol{x} = (x_1, x_2, \ldots, x_t)^T$ these transmissions. We also denote by a vector $\boldsymbol{b} = (b_1, b_2, \ldots, b_m)^T$ all the messages and by a matrix $A = (a_{l,j}) \in \mathbb{F}_q^{t \times m}$ all the coding coefficients. Then, we can write the encoding process $\phi_0 : \mathbb{F}_q^m \to \mathbb{F}_q^t$ in a matrix form: $\boldsymbol{x} = \phi_0(\boldsymbol{b}) = A\boldsymbol{b}$.

For client $i \in [n]$, the decoding process is a mapping $\phi_i : \mathbb{F}_q^t \times \mathbb{F}_q^{|S_i|} \to \mathbb{F}_q^{[|R_i|]} \times 2^{R_i}$, where $\mathbb{F}_q^{[|R_i|]}$ denotes the set $\emptyset \cup \mathbb{F}_q^1 \cup \mathbb{F}_q^2 \ldots \cup \mathbb{F}_q^{|R_i|}$ and $2^{R_i}$ denotes the power set of $R_i$. The decoding function takes the broadcast transmissions $\boldsymbol{x}$ and the side information $\{b_j\}_{j \in S_i}$ as input and output the new messages with indices $D_i$ that client $i$ can decode, i.e., $(\{b_j\}_{j \in D_i}, D_i) = \phi_i(\boldsymbol{x}, \{b_j\}_{j \in S_i})$.

The coding coefficient matrix $A$ that the server creates is perfectly known by all clients when decoding. This is a common assumption used in index coding and pliable index coding literature [5]-[9], [12], [13].

Let us denote by $\phi = (\phi_0, \phi_1, \ldots, \phi_n)$ the broadcast transmission scheme, i.e., encoding and decoding, for the server and all clients.

We will use the following decoding criterion directly from [12] that determines whether client $i$ is able to decode some new message $j \in R_i$, given a certain transmission scheme.

**Lemma 1** (Decoding Criterion, Lemma 1 in [12])**.** *Given the transmission scheme with the coding coefficient matrix $A$, client $i$ can decode a message $j \in R_i$, if and only if*

$$\boldsymbol{a}_j \notin span\{\boldsymbol{a}_{j'} | j' \in R_i \backslash \{j\}\},$$

*where $\boldsymbol{a}_j$ is the $j$-th column vector of $A$ corresponding to message $j$, and $span\{\boldsymbol{a}_{j'} | j' \in R_i \backslash \{j\}\}$ denotes the linear space spanned by the column vectors of $A$ corresponding to messages in $R_i$ other than $j$.*

If $A \in \mathbb{F}_q^{1 \times m}$, then the broadcast transmission can be expressed as $a_1 b_1 + a_2 b_2 + \ldots + a_m b_m$. We then have the "one hot" criterion to determine whether one transmission can satisfy client $i$: if and only if in the set $\{a_j | j \in R_i\}$, there are exactly one non-zero value and all the others are $0$.

**Preference model:** To formally describe the preferences of clients, we first define 3 types of benefit as follows.

- *Benefit.*

– *Benefit of message $b_j$ for client $i$.* This is the benefit a message $b_j$ brings to client $i$ when client $i$ only receives message $b_j$. We denote this by $s_i(b_j) \geq 0$, or $s_i(j)$ for short.

– *Benefit of a set of messages $\{b_j\}_{j \in D}$ or a transmission scheme $\phi$ for client $i$.* If client $i \in [n]$ can successfully decode a set of messages $\{b_j\}_{j \in D}$, where $D$ is the set of indices, then the benefit for client $i$, $s_i(\{b_j\}_{j \in D})$ or $s_i(D)$ for short, is defined as $s_i(D) = \max_{j \in D}\{s_i(j)\}$ for $D \neq \emptyset$, and $s_i(D) = 0$ for $D = \emptyset$, according to the *maximum benefit rule*[2]. If a transmission scheme $\phi$ enables client $i$ to decode the set of messages indexed by $D_i$, then we also say that the benefit this transmission scheme $\phi$ brings to client $i$ is $s_i(\phi) \triangleq s_i(D_i)$. When it is clear from the context, we will simply denote $s_i(\phi)$ or $s_i(D_i)$ as $s_i$.

— *Total benefit of a transmission scheme*. If a transmission scheme $\phi$ enables each client $i \in [n]$ to decode a set of messages indexed by $D_i$, then the total benefit across all clients is calculated as $B = \sum_{i \in [n]} s_i(\phi)$.

The three benefits we discussed above are distinguished by the notation $s_i(j)$, $s_i$, and $B$. Note that the theorems derived in Sections III-V refer to the total benefit $B$.

We emphasize that in this paper, we focus on the case where clients accrue benefits only if they receive a message they do not already have in their side information set. We discuss possible extensions of this model in Section VII.

● *Preference model*. According to the restrictions on benefits among messages for a client, we consider the following preference models in this paper.

— The *Borda count model* [10] assumes that for every client $i$, she ranks the $|R_i|$ messages according to her preference and then assigns individual benefits $|R_i|, |R_i| - 1, |R_i| - 2, \ldots, 1$ to $s_i(j)$ with respect to this ranking. In other words, the message ranked first gives benefit $|R_i|$, ranked second gives benefit $|R_i| - 1$, etc.

— The *bimodal benefit model* assumes that a fraction $F$ of the messages are much more desirable than the remaining $(1 - F)$ fraction for every client $i$. It is a modification of the Borda count model by scaling up the benefits of a fraction $F$ of the highest ranked messages by a factor $G$. In particular, for client $i$, we have the benefits $G|R_i|, G(|R_i| - 1), G(|R_i| - 2), \ldots, G(\lfloor (1 - F)|R_i| \rfloor + 1)$ for a $F$ fraction of the highest ranked messages and benefits $\lfloor (1 - F)|R_i| \rfloor, \lfloor (1 - F)|R_i| \rfloor - 1, \ldots, 1$ for a $1 - F$ fraction of the lowest ranked messages, where $\lfloor y \rfloor$ denotes the largest integer less than or equal to $y$. The parameter $G$ determines how separated (bimodal) the two sets of messages are.

— The *general benefit model* assigns an *arbitrary weight* to individual benefit of message $j$ for client $i$, without putting restrictions except the non-negativity, $s_i(j) = w_{ij} \geq 0$.

**Performance metrics:** We are interested in the tradeoff between the number of broadcast transmissions $t$ and the corresponding achievable benefit $B$.

---

[2]See Section VII for a discussion on another possible rule.

## B. Problem Formulations

Here, we first express in a unified notation the index coding and pliable index coding problem that have been examined in the literature before, and then introduce the new formulations we will examine through theoretical analysis in this paper. We show connections between our new problem formulations and the index coding problem, the pliable index coding problem, indicating that both index coding and pliable index coding problems are two special cases of our new formulations.

### Past Formulations

**Index Coding:** Each client requests a specific message that she does not have[3]; if client $i$ would like to receive $b_{j_i}$, we set $s_i(j) = 1$ for $j = j_i$ and $s_i(j) = 0$ for $j \in R_i \backslash \{j_i\}$. Thus $s_i$ takes values either 0 or 1, depending on whether client $i$ can decode $b_{j_i}$ or not, and $0 \leq B \leq n$. Index coding asks for the minimum number of transmissions to achieve the maximum possible benefit $B = n$, i.e., so that all clients receive the message they have requested.

*This problem is NP hard and requires in the worst case $\Omega(n)$ transmissions [6], [14], [7], and almost surely $\Theta(\frac{n}{\log(n)})$ transmissions for random graphs [15], [16].*

**Pliable Index Coding:** Each client is happy to receive any message she does not have (without any preference). We thus set $s_i(j) = 1$, for all $i$ and $j \in R_i$. Then $s_i$ takes value 1 or 0, depending on whether client $i$ can decode some message in $R_i$ or none in $R_i$. Therefore, we have $0 \leq B \leq n$. Pliable index coding asks for the minimum number of transmissions to achieve benefit $B = n$.

*This problem is NP hard, but there exist polynomial time algorithms that require in the worst case $O(\log^2 n)$ transmissions [12], [13].*

### New Formulations

The following formulations describe some scenarios for which we derive theoretical results. In each case, we ask what is the benefit $B$ with $t$ transmissions.

**P1. Borda count model with no side information:** No side information implies that $R_i = [m]$. We consider the Borda count model, where $s_i(j)$ defines for each client $i$ a permutation of $[m]$. Thus $0 \leq B \leq nm$.

---

[3]When a client requests $k$ specific messages, it is equivalent to the case that there are $k$ clients who have the same side information set and each requests one of the $k$ specific messages.

**P2. Borda count model with side information of equal size:** We assume that the size of the side information set $|S_i|$ is equal to $m - k$ for all clients, and thus $s_i(j)$ defines for client $i$ a permutation of $[k]$ over the remaining $k$ unknown messages. In this case $0 \leq B \leq nk$.

**P3. General benefit model with side information of arbitrarily size:** If the size of the side information set for each client is arbitrary, it is not reasonable to use Borda count model to calculate the benefit based on the ranking of messages in $R_i$, as it would give unfair weights to different clients. We assume instead that (fair) benefits $s_i(j) = w_{ij} \geq 0$ for $j \in R_i$ are provided as input.

*C. Problem Representation*

We can represent our problem instance using a $n \times m$ *benefit matrix* $\Pi$, where we have $n$ rows representing $n$ clients and $m$ columns representing $m$ messages. For the $(i, j)$ entry, if $j \in R_i$, we assign the value $s_i(j)$ representing the benefit client $i$ receives from a single message $j$; if $j \in S_i$, we assign a non defined value $x$. For example, the following matrices $\Pi_1$, $\Pi_2$, and $\Pi_3$ in eq. (1) represent the 3 types of aforementioned problems P1, P2, and P3, consisting of 3 clients and 5 messages. For problem $\Pi_1$, the benefits of messages for each client are a permutation of $[5]$; the single message $b_2$ gives client 3 the benefit $s_3(2) = 1$; and if client 2 receives $\{b_2, b_3, b_4\}$, she gets benefit $\max\{2, 5, 3\} = 5$. For problem $\Pi_2$, there are 2 side information messages for all 3 clients; the benefits of 3 unknown messages for each client are a permutation of $[3]$; and if the server makes a broadcast transmission $b_3 + b_4$, then this transmission enables client 1 to decode $b_3$, client 2 to decode $b_4$, and client 3 to decode none, resulting in a total benefit $2 + 3 = 5$. For problem $\Pi_3$, the size of the side information can be different for different clients, and non-negative benefits are assigned for $s_i(j)$.

$$\Pi_1 = \begin{bmatrix} 1 & 3 & 2 & 4 & 5 \\ 4 & 2 & 5 & 3 & 1 \\ 2 & 1 & 4 & 5 & 3 \end{bmatrix}, \quad \Pi_2 = \begin{bmatrix} 1 & 3 & 2 & x & x \\ x & 2 & x & 3 & 1 \\ 2 & 1 & x & x & 3 \end{bmatrix}, \quad \Pi_3 = \begin{bmatrix} 1 & x & 2 & x & x \\ x & 2 & x & 2 & 1 \\ 2 & 1 & 2.2 & x & 0 \end{bmatrix}. \quad (1)$$

## III. No Side Information (P1)

This is the simplest case we examine. This problem is close to the rank aggregation problems studied in the literature [10], [17], [18], [19], the difference being that only the highest ranked message a client receives counts towards the total benefit (i.e., the maximum benefit rule).

---

**Algorithm 1** Greedy algorithm for P1.

---

1: **Input**: benefit matrix $\Pi$ and number of columns to select $t$.
2: **Output**: a set of columns $\mathcal{T}$.
3: **Initialization**: set $\mathcal{T} = \emptyset$, $B_{\mathcal{T}} = 0$.
4: **for** $\tau = 1 : t$ **do**
5:    $j = \arg \max_{j' \in [m] \setminus \mathcal{T}} B_{\mathcal{T} \cup \{j'\}}$
    find a column $j$ to maximize the benefit given current $\tau - 1$ selected columns $\mathcal{T}$.
6:    $\mathcal{T} = \mathcal{T} \cup j$.
7: **end for**

---

Interestingly, while the rank aggregation problem is polynomial time using the Borda count model, taking into account only the highest score message makes the problem NP-hard.

We next describe our results, and provide the proofs in the Appendix A. Note that since clients have no side information in problem P1, and since all clients receive all transmissions, if one client can decode a message, all other clients can also do so - there is no differentiation between clients. Moreover, because $t$ broadcast transmissions can enable each client to decode at most $t$ messages, we can with no loss of generality use uncoded transmissions, there are no benefits in using a linear or nonlinear coding scheme and for example the bounds we derive apply even if we are allowed to use nonlinear schemes (see also Appendix A.1).

**The problem is NP-hard.**

**Theorem 1.** *The bandwidth aware recommendation problem using Borda count model with no side information (P1) is NP-hard.*

The proof uses a reduction from the set cover problem (see Appendix A.2).

**Greedy selection achieves an approximation ratio of** $1.58$**.** We use the benefit matrix $\Pi$ introduced in Section II to represent the problem instance. For $t = 1$, we simply need to find the column of the matrix $\Pi$, whose elements have the highest sum (this would be the benefit $B$). For $t > 1$, we need to select $t$ columns in a set $\mathcal{T}$ such that $B_{\mathcal{T}}$ is as large as possible, where we denote by $B_{\mathcal{T}}$ the benefit from the choice of a set of columns $\mathcal{T}$. Alg. 1 describes a straightforward greedy algorithm, that is sufficient to achieve a constant approximation ratio. Let $\mathcal{T}^*$ be the optimal selection of $t$ columns, and $B^* \triangleq B_{\mathcal{T}^*}$ the optimal benefit achieved by this selection over this problem instance. We have the following theorem.

**Theorem 2.** *For any P1 problem instance, given that the server can make $t$ broadcast trans-*

*missions, Alg. 1 can achieve an approximation ratio at least $1/(1-(1-\frac{1}{t})^t)$, namely,*

$$\frac{B_{\mathcal{T}}}{B^*} \geq 1 - (1-\frac{1}{t})^t \geq 1 - \frac{1}{e}. \tag{2}$$

The proof is given in Appendix A.3. From Theorem 2, we can see that for any $t$, the approximation ratio is bounded by a constant factor $1/(1-\frac{1}{e}) = 1.58$.

**Bounds on the optimal benefit $B^*$.**

**Theorem 3.** *The optimal benefit $B^*$ of problem P1 can be lower bounded by*

$$B^* \geq \frac{tn(m+1)}{t+1} \triangleq \mu. \tag{3}$$

*Moreover, if $n \geq 6t\log(m)$, there exist problem P1 instances such that*

$$B^* \leq (1+\delta)\frac{tn(m+1)}{t+1} = (1+\delta)\mu, \tag{4}$$

*where $\delta = \sqrt{\frac{6t\log(m)}{n}}$.*

The proof is in Appendix A.4. We underline that the upper bound does not apply for all instances (it is trivial to create instances where we get $B^* = nm$) but only for the worst case instances. Note that if $\delta$ is small, the lower and upper bounds have the same order of magnitude, which implies that the bounds become tight. Moreover, if the number of clients increases to $n > O(t^3\log(m))$, then the (worst case instances) upper bound can be simplified to $O(mn(1-\frac{1}{t}))$, which is close to the optimal $nm$. In particular, if we consider all possible $n = m!$ clients, each having distinct benefits among the messages (a permutation of $[m]$), then the benefit achieved by *any* $t$-selection is $\mu = \frac{tn(m+1)}{t+1}$.

**Average case analysis.** Here, we assume uniform distribution of a client's benefits among messages, and calculate the expected benefit. In particular, the benefit matrix $\Pi$ is generated as follows: for each row (a client's benefits among the messages), select uniformly and independently a permutation from all $m!$ permutations of $[m]$ (with replacement).

**Theorem 4.** *By uniformly and independently generating the benefit matrix $\Pi$, the optimal benefit of P1 averaged over the benefit matrix $\Pi$ satisfies:*

$$\mathbb{E}_\Pi B \geq \mu + n\Delta(m, n, t),$$

*where* $\mu = \frac{tn(m+1)}{t+1}$, $\Delta(2,n,1) = \frac{1}{\sqrt{2\pi n}}$ *for even* $n$, $\Delta(2,n,1) = \frac{1}{\sqrt{2\pi(n-1)}}$ *for odd* $n \geq 3$, $\Delta(m,n,t) \approx \frac{1}{\sqrt{2\pi n}}\sigma(m,t)$ *for large* $n$, *and* $\sigma(m,t) = \sqrt{\frac{(m+1)(m-t)t}{(t+1)^2(t+2)}}$ *is the standard deviation of a client's benefit when we randomly select* $t$ *columns.*

Note that we have already shown in Theorem 3 that the worst case benefit is $\mu$; the above theorem shows that the average is higher by at least $n\Delta(m,n,t)$. The proof is provided in Appendix A.5.

**Illustrating example.** Consider an instance with $n = 5$ clients, $m = 4$ messages and the $5 \times 4$ benefit matrix

$$\Pi = \begin{bmatrix} 4 & 3 & 1 & 2 \\ 1 & 4 & 3 & 2 \\ 1 & 3 & 4 & 2 \\ 1 & 3 & 2 & 4 \\ 1 & 4 & 3 & 2 \end{bmatrix}. \tag{5}$$

We can see that, to serve to all clients their first preference (benefit $B = 20$) we need $t = 4$ transmissions; yet with only $t = 1$ transmission (second column) we can already serve to all clients either their first or their second preference (benefit $B = 17$). Moreover, in a situation where the server recommends to send to each client their first preference but only one of these messages is delivered (in time) because of bandwidth constraints, we would in the worst case achieve benefit $B = 8$ (e.g., only the first message is delivered); thus taking into account the bandwidth constraints can more than double the benefit. This difference can be magnified proportionally to a parameter $G$, if instead of Borda count we used bimodal benefit model with gain factor $G$.

## IV. Side Information of Equal Size (P2)

We now look at the case where all clients have side information of the same size $|S_i| = m - k$ and thus $|R_i| = k$, $\forall\, i$. We again assume the Borda count model. In this case, because of the difference in side information for different clients, the same broadcast transmissions enable different clients to decode different messages. Therefore, coding will help to improve the efficiency.

**Bounds on the optimal benefit:** We are interested in the optimal benefit $B^*$ we can achieve with $t$ transmissions (recall that $0 \leq B \leq kn$). We next prove a lower bound on the performance of the optimal benefit through dynamic programming. This bound is achieved through linear coding schemes.

**Theorem 5.** *The optimal benefit $B^*$ satisfies*

$$B^* \geq \begin{cases} \frac{kn}{4e} & \text{for} \quad t = 1, \\[2mm] \frac{kn}{e}\left(\frac{1}{2} - \frac{1}{8e} + \frac{1}{16e^2}\right) & \text{for} \quad t = 2, \\[2mm] \frac{kn}{e}\left(\frac{3}{4} - \frac{1}{4e} + \frac{1}{16e^2}\right) & \text{for} \quad t = 3, \\[2mm] \frac{kn}{e}\left(1 - \frac{5}{8e} + \frac{13}{64e^2}\right) & \text{for} \quad t = 4, \\[2mm] kn\left(1 - \frac{4e}{t} + \frac{12e}{t^2}\right) & \text{for} \quad 5 \leq t < k \\[2mm] kn & \text{for} \quad t = k. \end{cases} \tag{6}$$

The proof of this theorem is constructive: we provide a randomized algorithm and show that it achieves on average the performance prescribed in the theorem, which implies that the optimal performance can only be better. The approximation ratio of this scheme is $O(1)$, as the best achievable benefit is $kn$. We also note that if $t = k$, we can easily achieve $B = kn$: the server can use a Maximum Distance Separable (MDS) erasure correcting code to create $k$ linear combinations to transmit, so that each client using her side information can solve for the $k$ messages she misses. We next show the proof of the theorem.

• For $t = 1$, assume that the server makes the transmission $x_1 = a_1 b_1 + a_2 b_2 + \ldots a_m b_m$ where $a_j$, $j \in [m]$, are the constant coding coefficients (we will call the vector $\boldsymbol{a} = (a_1, a_2, \ldots, a_m)$ the row coding vector). We use the strategy that selects i.i.d. random values for the coding coefficients, by setting $a_j = 1$ with probability $1/k$, and $a_j = 0$ otherwise.

*Claim 1:* There exists a binary row coding vector $\boldsymbol{a}^\xi$ that enables at least $\frac{n\xi}{ke}$ clients to decode a message in their request set with individual benefit more than $k - \xi$, for $\xi = 1, 2, \ldots, k$, and thus to achieve a total benefit of at least $\frac{(k+1-\xi)n\xi}{ke}$ for these clients.

*Proof.* Without loss of generality, assume that client $i$ has the request set $R_i = \{b_{l_1}, b_{l_2}, \ldots, b_{l_k}\}$ with individual benefits $s_i(l_1) = k$, $s_i(l_2) = k-1, \ldots, s_i(l_k) = 1$. Client $i$ can decode a message with benefit at least $k - \xi + 1$, i.e., can decode some $b_{l_j}$ with $j \leq \xi$, if and only if $a_{l_j} = 1$ and $a_{l_1} = a_{l_2} = \ldots = a_{l_{j-1}} = a_{l_{j+1}} = \ldots = a_{l_k} = 0$, according to the "one hot" criterion derived from Lemma 1. Indeed, we can then express the server transmission as $x_1 = b_{l_j} + \sum_{l \in S_i} a_l b_l$; client $i$ can remove from $x_1$ the part $\sum_{l \in S_i} a_l b_l$ using her side information, and decode $b_{l_j}$. The probability that such an event happens is:

$$\binom{\xi}{1} \frac{1}{k} (1 - \frac{1}{k})^{k-1} \le \frac{\xi}{ke} \triangleq p_\xi. \tag{7}$$

Hence, randomly selecting a row coding vector would enable on average at least $np_\xi = \frac{n\xi}{ke}$ clients to decode messages of individual benefit at least $k - \xi + 1$, and thus, from the averaging principle, there exists at least one coding vector $\boldsymbol{a}^\xi$ that also enables this. $\qquad\square$

• For $t > 1$, we consider the following dynamic programming problem. We consider $t$ stages, each corresponding to one transmission. At stage $\tau$, $1 \le \tau \le t$, the server can select one of $k$ actions, which of the $k$ possible $\boldsymbol{a}^\xi$ vectors, for $\xi = 1, 2, \ldots, k$, to use. In particular, we proceed as follows:

– At the beginning of stage 1, no transmission has yet been made and there are $n_1 = n$ clients in the system. The server chooses an action $\xi_1 \in [k]$, i.e., uses the row coding vector $\boldsymbol{a}^{\xi_1}$ to make a transmission. From Claim 1, this transmission enables $\frac{n_1 \xi_1}{ke}$ clients to decode a message that gives a benefit at least $k + 1 - \xi_1$, and thus, we can achieve a total benefit $B_1 \ge \frac{(k+1-\xi_1)n_1\xi_1}{ke}$ for these $\frac{n_1 \xi_1}{ke}$ clients. We then remove these $\frac{n_1 \xi_1}{ke}$ clients from the system and denote the remaining number of clients by $n_2 = n_1(1 - \frac{\xi_1}{ke})$.

– At the beginning of stage 2, we only consider the $n_2$ clients; similarly to before, the server chooses an action $\xi_2 \in [k]$ to enable $\frac{n_2 \xi_2}{ke}$ clients decode a message that gives a benefit at least $k + 1 - \xi_2$. At this point we have achieved a total benefit $B_2 \ge B_1 + \frac{(k+1-\xi_2)n_2\xi_2}{ke}$, where the first term $B_1$ is achieved at stage 1 and the second term is the total benefit achieved for these $\frac{n_2 \xi_2}{ke}$ clients at stage 2. We remove these $\frac{n_2 \xi_2}{ke}$ clients from the system and denote the remaining number of clients by $n_3 = n_2(1 - \frac{\xi_2}{ke})$.

– Continuing along the same lines, at the beginning of stage $\tau = 3, 4, \ldots, t$, we have $n_\tau$ clients to consider; the server chooses an action $\xi_\tau$ that enables to achieve benefit $B_\tau \ge B_{\tau-1} + \frac{(k+1-\xi_\tau)n_\tau\xi_\tau}{ke}$; and we set $n_{\tau+1} = n_\tau(1 - \frac{\xi_\tau}{ke})$.

Let $J_\tau(n_\tau)$ be the benefit the $n_\tau$ clients can receive for the *remaining* $t + 1 - \tau$ stages. We have the following Bellman equation:

$$J_\tau(n_\tau) = \max_{\xi_\tau \in [k]}\{(k+1-\xi_\tau)\frac{\xi_\tau n_\tau}{ke} + J_{\tau+1}(n_\tau(1-\frac{\xi_\tau}{ke}))\}, \tag{8}$$

with $J_{t+1}(n_{t+1}) = 0$ (as we will only make $t$ transmissions).

From the above equation, we can see that the benefit achieved using this scheme is $B = J_1(n)$.

- If $t = 1$, we set $\xi_1 = \frac{k}{2}$ ($k$ even) and $\xi_1 = \frac{k+1}{2}$ ($k$ odd), and get $J_1(n) \geq \frac{kn}{4e}$.

- If $t = 2$, we set $\xi_1 = \lceil \frac{k}{2}(1-\frac{1}{4e}) \rceil$ and $\xi_2 = \lceil \frac{k}{2} \rceil$, and get $J_1(n) \geq \frac{kn}{e}(\frac{1}{2}-\frac{1}{8e}+\frac{1}{16e^2})$.

- If $t = 3$, we set $\xi_1 = \lceil \frac{k}{2}(1-\frac{1}{2e}) \rceil$, $\xi_2 = \lceil \frac{k}{2}(1-\frac{1}{4e}) \rceil$ and $\xi_3 = \lceil \frac{k}{2} \rceil$, and get $J_1(n) \geq \frac{kn}{e}(\frac{3}{4}-\frac{1}{4e}+\frac{1}{16e^2})$.

- If $t = 4$, we set $\xi_1 = \lceil \frac{k}{2}(1-\frac{3}{4e}) \rceil$, $\xi_2 = \lceil \frac{k}{2}(1-\frac{1}{2e}) \rceil$, $\xi_3 = \lceil \frac{k}{2}(1-\frac{1}{4e}) \rceil$ and $\xi_4 = \lceil \frac{k}{2} \rceil$, and get $J_1(n) \geq \frac{kn}{e}(1-\frac{5}{8e}+\frac{13}{64e^2})$.

- For $t > 4$, we have the following claim.

*Claim 2:* $J_\tau(n) \geq kn(1 - \frac{4e}{t+1-\tau} + \frac{12e}{(t+1-\tau)^2})$ for $t > 4$.

*Proof.* We use the backward induction method for the last 5 stages. By setting $\xi_{t-4} = \lceil \frac{k}{2}(1-\frac{1}{e}+\frac{5}{8e^2}) \rceil$, $\xi_{t-3} = \lceil \frac{k}{2}(1-\frac{3}{4e}) \rceil$, $\xi_{t-2} = \lceil \frac{k}{2}(1-\frac{1}{2e}) \rceil$, $\xi_{t-1} = \lceil \frac{k}{2}(1-\frac{1}{4e}) \rceil$ and $\xi_t = \lceil \frac{k}{2} \rceil$, we can get $J_{t-4}(n_{t-4}) \geq \frac{kn_{t-4}}{e}(\frac{5}{4}-\frac{9}{8e}+\frac{39}{64e^2}) \doteq 0.33kn_{t-4}$. Therefore, we have the initial condition:

$$J_{t-4}(n_{t-4}) \doteq 0.33kn_{t-4} \geq kn_{t-4}(1-\frac{4e}{t+1-(t-4)}+\frac{12e}{(t+1-(t-4))^2}) \doteq 0.13kn_{t-4}.$$

Assume that $J_{\tau+1}(n_{\tau+1}) \geq kn_{\tau+1}(1-\frac{4e}{t+1-(\tau+1)}+\frac{12e}{(t+1-(\tau+1))^2})$ holds for $\tau+1$, then consider $J_\tau(n_\tau)$ ($\tau < t-4$):

$$\begin{aligned} J_\tau(n_\tau) &= \max_{\xi_\tau \in [k]}\{(k+1-\xi_\tau)\frac{\xi_\tau n_\tau}{ke} + J_{\tau+1}(n_\tau(1-\frac{\xi_\tau}{ke}))\} \\ &\geq \max_{\xi_\tau \in [k]}\{(k+1-\xi_\tau)\frac{\xi_\tau n_\tau}{ke} + kn_\tau(1-\frac{\xi_\tau}{ke})(1-\frac{4e}{t+1-(\tau+1)}+\frac{12e}{(t+1-(\tau-1))^2})\} \\ &\geq kn_\tau(1-\frac{4e}{t+1-\tau}+\frac{12e}{(t+1-\tau)^2}), \end{aligned}$$

where the first inequality holds due to the hypothesis and the property of the Bellman equation; the second inequality holds by setting $\xi_\tau$ to be an integer between $\xi' = \frac{2ke}{t-\tau} - \frac{6ke}{(t-\tau)^2} + \frac{1}{2}$ and $\xi'' = \frac{2ke}{t-\tau} - \frac{6ke}{(t-\tau)^2} - \frac{1}{2}$. If we define $f(\xi) = (k+1-\xi)\frac{\xi n_\tau}{ke} + kn_\tau(1-\frac{\xi}{ke})(1-\frac{4e}{t+1-(\tau+1)}+\frac{12e}{(t+1-(\tau+1))^2})$, then we have $f(\xi_\tau) \geq \min\{f(\xi'), f(\xi'')\} \geq kn_\tau(1-\frac{4e}{t+1-\tau}+\frac{12e}{(t+1-\tau)^2})$. Therefore, claim 2 holds. By setting $\tau = 1$, we get $J_1(n) \geq kn(1-\frac{4e}{t}+\frac{12e}{t^2})$. $\qquad \square$

**Algorithm 2:** From the constructive proof of Theorem 5, we propose a dynamic programming based algorithm to solve problem P2, as shown in Alg. 2. This proposed algorithm operates

---

**Algorithm 2** Dynamic programming algorithm for solving P2.

---

1: **Input**: number of messages $m$, number of clients $n$, request sets $R_i, \forall i \in [n]$, benefit matrix $\Pi$, size of request set $k$, and number of transmissions $t$.
2: **Output**: coding matrix $A \in \{0, 1\}^{t \times m}$.
3: **Initialization**: set the client set $\mathcal{N} = [n]$;
4: **for** $\tau = 1 : t$ **do**
5:     **if** $\tau = t$ **then**
6:         Set the action $\xi_\tau = \lceil \frac{k}{2} \rceil$.
7:     **else if** $\tau = t - 1$ **then**
8:         Set the action $\xi_\tau = \lceil \frac{k}{2}(1 - \frac{1}{4e}) \rceil$.
9:     **else if** $\tau = t - 2$ **then**
10:        Set the action $\xi_\tau = \lceil \frac{k}{2}(1 - \frac{1}{2e}) \rceil$.
11:     **else if** $\tau = t - 3$ **then**
12:        Set the action $\xi_\tau = \lceil \frac{k}{2}(1 - \frac{3}{4e}) \rceil$.
13:     **else if** $\tau = t - 4$ **then**
14:        Set the action $\xi_\tau = \lceil \frac{k}{2}(1 - \frac{1}{e} + \frac{5}{8e^2}) \rceil$.
15:     **else**
16:        Set the action $\xi_\tau = \lceil \frac{2ke}{t-\tau} - \frac{6ke}{(t-\tau)^2} - \frac{1}{2} \rceil$.
17:     **end if**
18:     Find a row coding vector $\boldsymbol{a}^{\xi_\tau}$ as the $\tau$-th row of $\boldsymbol{A}$ with respect to the action $\xi_\tau$ and clients $\mathcal{N}$ using Alg. 4, the derandomization function.
19:     Remove all $i$ from $\mathcal{N}$, if $i \in \mathcal{N}$ receives the benefit no less than $k + 1 - \xi_\tau$, given coding vector $\boldsymbol{a}^{\xi_\tau}$.
20: **end for**

---

in stages and at each stage selects what row coding vector to use for a transmission so as to satisfy a certain fraction of clients. In the proof of Theorem 5, we use a random generation argument to show the existence of the desired actions in Claim 1. However, we can easily derandomize it using a deterministic algorithm in polynomial time: we sequentially visit the entries of the row coding vector and decide whether to assign a value $0$ or $1$ depending on how the benefit would increase, as described in detail in Alg. 4 in Appendix B.

**Benefits of Coding:** As is the case in index coding, leveraging side information enables to use coding and convey through the same transmission different messages to clients. We next compare, over two sets of instances, the ratio of the benefit we get using side information to that without using side information (i.e., uncoded transmissions).

- *Ratio of* $\frac{n}{t}$. Assume that for each pair of clients $i_1$ and $i_2$, the request sets $R_{i_1}$ and $R_{i_2}$ do not overlap, i.e., $R_{i_1} \cap R_{i_2} = \emptyset$ for all $i_1 \neq i_2$. Assume that each client receives a maximum benefit of $s$ if she can decode her most preferred message. In this case, the best uncoded $t$ transmissions are to choose the $t$ messages such that $t$ clients receive the maximum benefits (and others without benefits), achieving a total benefit of $ts$. With $t$ encoded transmissions each client can decode her most preferred message. Therefore, the ratio is $\frac{ns}{ts} = \frac{n}{t}$.

- *Ratio of* $\frac{2n}{t(k+1)}$. Consider an instance with $m$ messages and $n = m$ clients. All clients have a request set of the same size, i.e., $|R_i| = k < t$, $\forall i$. The clients are partitioned into groups: for any two clients $i_1$ and $i_2$ in different groups, their request sets $R_{i_1}$ and $R_{i_2}$ do not overlap, i.e., $R_{i_1} \cap R_{i_2} = \emptyset$; for any two clients $i_1$ and $i_2$ in the same group, their request sets $R_{i_1}$ and $R_{i_2}$ are the same, i.e., $R_{i_1} = R_{i_2}$. In each group, the number of clients is equal to the cardinality of the request set $k$, and thus we have $k$ clients requiring $k$ messages. We assign the associated benefit submatrix of each group to be a *Latin square*, i.e., each requested message gives different individual benefits for these $k$ clients, from $1$ to $k$. Hence, the benefit submatrix has no same elements in the same row and the same column. Therefore, for the uncoded $t$ transmissions, this instance will give a total benefit of $\frac{k(k+1)t}{2}$. For the coded $t$ transmissions, this instance will give a total benefit of $nk$, when we use MDS coding scheme to send all the missing messages. Hence, the ratio is $\frac{2n}{t(k+1)}$.

## V. SIDE INFORMATION OF ARBITRARY SIZE (P3)

Here, we do not put restrictions on the non-negative individual benefit $s_i(j) = w_{ij} \geq 0$ that message $j$ provides for client $i$, and make no assumptions on the size of the side information set. This is the most general case that admits P1 and P2 as special cases. We solve this problem using a mapping to the MWIS problem[4]. The mapping and the bounds we derive apply for linear coding schemes.

We consider the cases $t = 1$ and $t > 1$ separately. For $t = 1$, we show that the problem P3 is polynomial time reducible to the MWIS problem; while for $t > 1$, we show that the problem P3 is equivalent to an MWIS problem, yet the reduction maybe exponential. Leveraging polynomial time reduction for the $t = 1$ case, we propose a heuristic algorithm to solve problem P3 by repeatedly using an MWIS solver to solve a $t = 1$ sub-problem.

### A. Mapping to the MWIS Problem

**Problem P3 with** $t = 1$**:** Assume that the server uses a binary row coding vector $\boldsymbol{a} = (a_1, \ldots, a_m)$ to make a transmission $x_1 = a_1 b_1 + \ldots + a_m b_m = b_{j_1} + b_{j_2} + \ldots + b_{j_l}$, where the indices $j_1, j_2, \ldots, j_l$ correspond to the non-zero coding coefficients.

Recall from Lemma 1 that client $i$ can decode the message $b_{j_1}$ from $x_1$ if and only if this is the only message appearing in $x_1$ that she does not have; that is, $b_{j_1}$ belongs in her request set, $j_1 \in R_i$, and she has already as side information the rest of the messages appearing in $x_1$, $j_2, \ldots, j_l \in S_i$. Consider now the $|R_i|$ positions in the coding vector $\boldsymbol{a}$ that correspond to the $|R_i|$ messages client $i$ does not have. There are $|R_i|$ possible choices of coding coefficients for these positions, so that client $i$ can decode one of these message: making exactly one of these coefficients one, and the remaining $|R_i| - 1$ zero.

---

[4]The MWIS problem is the weighted version of the maximum independent set problem. It aims to find a set of vertices that have the maximum weighted sum in a given graph. For details of the problem, see, for example, [20].

If we depict these coefficients sequentially, the $|R_i|$ possible choices are $(1, 0, 0, \ldots, 0)_i$, $(0, 1, 0, \ldots, 0)_i$, …, $(0, \ldots, 0, 1)_i$, where we use the subscript $i$ to express that these correspond to the messages in $R_i$. Client $i$ can decode the first message in $R_i$ under the first choice, the second message under the second choice, etc. We call these $|R_i|$ possible choices the *partial assignments* for client $i$. Note that, here, for each client, we only consider the partial assignments that enable this client to decode a new message in $R_i$, hence, there are $|R_i|$ partial assignments for each client. For each partial assignment, we only care about the partial row coding vector corresponding to the messages in $R_i$.

We map each of the $|R_i|$ partial assignments for client $i$, for all $i \in [n]$, to a vertex in the MWIS instance; thus in total we create $\sum_{i \in [n]} |R_i|$ vertices. We assign weight $w_{ij}$ to the vertex corresponding to the partial assignment that enables client $i$ to decode message $j$[5]. We connect two vertices with an edge if the corresponding partial assignments cause conflict with each other: that is, there exists at least one common message to which one vertex assigns coefficient 0 and the other coefficient 1. Vertices corresponding to partial assignments of the same client $i$ are pair-wise connected, forming a clique, since these partial assignments are mutually exclusive. Vertices corresponding to partial assignments of different clients may be connected or not.

Given that each vertex of this graph specifies part of a row coding vector, an independent set specifies (perhaps in part) a feasible row coding vector that enables all clients with a vertex in this independent set to decode a message. Because we list all possible partial assignment choices for a client to decode a message, finding an MWIS enables to construct a row coding vector that leads to the maximum benefit. It is not hard to see that this mapping process from a P3 problem to an MWIS weighted graph instance is polynomial time: there are $\sum_{i \in [n]} |R_i|$ vertices and at most $\binom{\sum_{i \in [n]} |R_i|}{2}$ edges.

---

[5]If some weight $w_{ij} = 0$, then we can remove this vertex from the graph without losing total benefit.

**An example:** For example, we consider the problem instance $\Pi_3$ in eq. (1) that is reiterated as follows.

$$\Pi_3 = \begin{bmatrix} 1 & x & 2 & x & x \\ x & 2 & x & 2 & 1 \\ 2 & 1 & 2.2 & x & 0 \end{bmatrix}. \tag{9}$$

In this problem instance, we have $R_1 = \{1, 3\}$, $R_2 = \{2, 4, 5\}$, and $R_3 = \{1, 2, 3, 5\}$. There are two partial assignments that can satisfy client 1: $(1, 0)_1$, $(0, 1)_1$; three partial assignments for client 2: $(1, 0, 0)_2$, $(0, 1, 0)_2$, $(0, 0, 1)_2$; and four partial assignments for client 3: $(1, 0, 0, 0)_3$, $(0, 1, 0, 0)_3$, $(0, 0, 1, 0)_3$, $(0, 0, 0, 1)_3$. Recall that we depict coefficients in each $R_i$ sequentially, for instance the partial assignment $(1, 0)_1$ for client 1 specifies the values at positions 1 and 3 of the coding vector, i.e., would be compatible with any coding vector of the form $(1, *, 0, *, *)$ where $*$ denotes an arbitrary value.

We label these partial assignments as vertices $1, 2, \ldots, 9$ and show them in Fig. 1 (a). In each row, we list a partial assignment for some client and map it to a vertex in the graph, with the vertex number in the 6-th column and the weight in the last column. The weight is assigned according to which message can be decoded by this client using the partial assignment. The constructed graph is given in the adjacency matrix form in Fig. 1 (b), based on whether there is a conflict between two partial assignments or vertices. For example, partial assignments 2 and 7 have a conflict, since partial assignment 2 requires $a_3 = 1$ but partial assignment 7 requires $a_3 = 0$. Similarly, there is no conflict for partial assignments 4 and 8.

**Problem P3 with $t > 1$:** Using a similar approach, we can reduce P3 to the MWIS problem even when $t > 1$, at the cost of higher (in some cases exponential) complexity. The idea is that, instead of having subvectors of size $1 \times |R_i|$ as vertices, we now have submatrices of size $t \times |R_i|$ as vertices. The mapping proceeds as follows.

• List all possible partial assignments of $t$-dimensional vectors for each client $i$. Consider only the partial assignments that correspond to messages in $R_i$, such that some message in $R_i$ can be decoded by client $i$. Note that each partial assignment consists of $|R_i|$ $t$-dimensional vectors, and that we can list all the partial assignments that enable client $i$ to decode some message in $R_i$ according to the decoding criterion in Lemma 1. We map each partial assignment (for all clients) to vertices and assign to a partial assignment the weight $\max_{j \in D_i} s_i(j)$ according to the

Partial assignments  Adjacency matrix

| $a_1$ | $a_2$ | $a_3$ | $a_4$ | $a_5$ | $vertex$ | $w_{ij}$ | |
|---|---|---|---|---|---|---|---|
| 1 |  | 0 |  |  | 1 | 1 | Client 1 |
| 0 |  | 1 |  |  | 2 | 2 | Client 1 |
|  | 1 |  | 0 | 0 | 3 | 2 | Client 2 |
|  | 0 |  | 1 | 0 | 4 | 2 | Client 2 |
|  | 0 |  | 0 | 1 | 5 | 1 | Client 2 |
| 1 | 0 | 0 |  | 0 | 6 | 2 | Client 3 |
| 0 | 1 | 0 |  | 0 | 7 | 1 | Client 3 |
| 0 | 0 | 1 |  | 0 | 8 | 2.2 | Client 3 |
| 0 | 0 | 0 |  | 1 | 9 | 0 | Client 3 |

| | 1 | 2 | 3 | 4 | 5 | 6 | 7 | 8 | 9 |
|---|---|---|---|---|---|---|---|---|---|
| 1 | 0 | 1 | 0 | 0 | 0 | 0 | 1 | 1 | 1 |
| 2 | 1 | 0 | 0 | 0 | 0 | 1 | 1 | 0 | 1 |
| 3 | 0 | 0 | 0 | 1 | 1 | 1 | 0 | 1 | 1 |
| 4 | 0 | 0 | 1 | 0 | 1 | 0 | 1 | 0 | 1 |
| 5 | 0 | 0 | 1 | 1 | 0 | 1 | 1 | 1 | 0 |
| 6 | 0 | 1 | 1 | 0 | 1 | 0 | 1 | 1 | 1 |
| 7 | 1 | 1 | 0 | 1 | 1 | 1 | 0 | 1 | 1 |
| 8 | 1 | 0 | 1 | 0 | 1 | 1 | 1 | 0 | 1 |
| 9 | 1 | 1 | 1 | 1 | 0 | 1 | 1 | 1 | 0 |

(a)  (b)

Fig. 1: An example of mapping from P3 to MWIS for $t = 1$. (a) The first five columns correspond to the $5$ messages. In each row, we list a partial assignment for some client and map it to a vertex in the graph, with the vertex number in the 6-th column and the weight in the last column. Note that we only consider the partial assignment corresponding to messages in $R_i$ and give the weight based on which message can be decoded according to this partial assignment. (b) We construct the graph based on whether two partial assignments have a conflict or not. For example, partial assignments $2$ and $7$ have a conflict, since partial assignment $2$ requires $a_3 = 1$ but partial assignment $7$ requires $a_3 = 0$. Similarly, there is no conflict for partial assignments $4$ and $8$.

maximum benefit rule, where $D_i$ denotes the indices of messages that client $i$ can decode under this partial assignment.

• Connect with an edge two vertices/partial assignments that are conflict, namely, there exists at least one common message to which one vertex assigns a $t$-dimensional vector $\boldsymbol{a}_j$ and the other assigns a $t$-dimensional vector $\boldsymbol{a}'_j \neq \boldsymbol{a}_j$.

Given that each vertex of this graph specifies parts of a coding coefficient matrix, an independent set specifies (at least in part) a feasible coding coefficient matrix that enables all clients with a vertex in this independent set to decode a message. Because we list all possible partial assignment choices for a client to decode a message, finding an MWIS enables to construct a coding coefficient matrix that leads to the maximum benefit.

The complexity of this mapping depends on the values of $|R_i|$ and $t$. If $|R_i|$ and $t$ are of the order $\Omega(1)$, then we can list all possible partial assignments in polynomial time, thus we can perform a polynomial time reduction. However, this mapping is not in general polynomial, as it is possible that the number of assignments for a client with a large $|R_i|$ (for example, $|R_i| = m/2$) increases exponentially, even for $t = 2$. For example, consider that $t = 2$ and $|R_i| = m/2$, then for client $i$ to decode some message $j \in R_i$, there are at least $2^{m/2-1}$ partial assignments that are suitable. Indeed, we can assign the column coding coefficient vector $(1, 0)^T$ to message $j$, and choose either $(0, 1)^T$ or $(0, 0)^T$ as the column coding coefficient vector for the remaining $m/2 - 1$ messages.

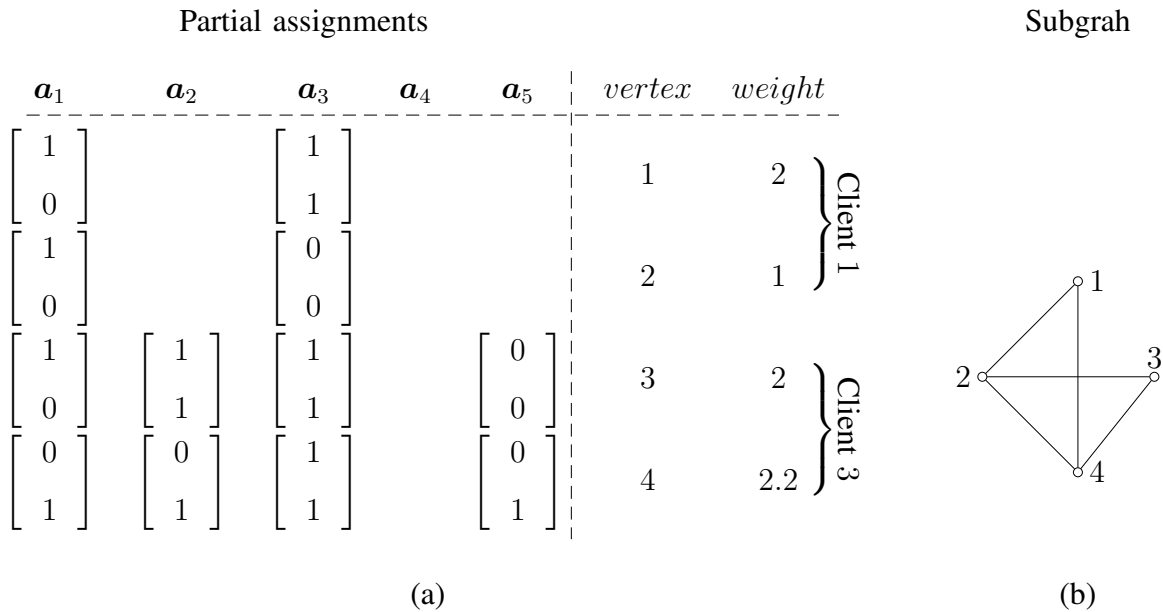

(a)            (b)

Fig. 2: An example of mapping from P3 to MWIS for $t = 2$ (part of the graph). (a) We list four of the partial assignments corresponding to four vertices in the graph: the first two are for client 1 and the last two are for client 3. Each partial assignment consists of $|R_i|$ 2-dimensional column coding coefficient vectors corresponding to messages in $R_i$. The weight depends on which messages the partial assignment enables client $i$ to decode. For example, the weight of vertex 1 is $\max\{1, 2\} = 2$, since the partial assignment 1 enables client 1 to decode both message 1 and 3. (b) We show part of the constructed graph based on whether two partial assignments have a conflict or not. For example, partial assignments 1 and 3 do not have a conflict, but partial assignments 1 and 4 have, since partial assignment 1 requires $\boldsymbol{a}_1 = (1, 0)^T$ but partial assignment 4 requires $\boldsymbol{a}_1 = (1, 1)^T$.

As a simple example, let us again consider the problem instance $\Pi_3$ in eq. (1). We do not list all possible partial assignments, but just list four partial assignments to illustrate the idea of mapping, as shown in Fig. 2.

*B. Polynomial Time Approximation Heuristics*

**Algorithm for t=1:** Given the MWIS connection, we can now translate any of the MWIS solvers to an algorithm for our problem when $t = 1$. As an example, the following theorem presents the achievable benefit using the MWIS polynomial time approximation algorithm in [20].

**Theorem 6.** *For problem P3, with $t = 1$ transmission, we can achieve a benefit of at least $\frac{W}{2(d_1-1)d_2+1}$ in polynomial time, where $W = \sum_{(i,j):j \in R_i} w_{ij}$ is the total weight of the instance, $d_1 = \max_{i \in [n]}\{|R_i|\} \leq m$ is the maximum size of side information sets among all clients, and $d_2 \leq n$ is the maximum number of request sets a message can belong to.*

*Proof.* The proof follows by observing that the maximum degree of each vertex in the graph is at most $2(d_1 - 1)d_2$, and then applying Theorem 3.4 in [20].

Indeed, consider a vertex $v$ that enables client $i$ to decode message $j \in R_i$. This vertex is connected to the remaining $|R_i| - 1$ vertices of the same client, which contributes to $v$ degree at most $d_1 - 1$. Now consider another client $i' \neq i$. If $j \in R_{i'}$, then only one of the partial assignments for client $i'$ does not have a conflict with $v$, the one that enables client $i'$ to decode $j$. Thus counting each $i' \neq i$ we may have additional degree of at most $(d_2 - 1)(d_1 - 1)$. Finally, consider $j' \in R_i \cap R_{i'}$ for some $j' \neq j$. In this case, the vertex $v'$ that enables client $i'$ to decode message $j'$ will be connected to $v$, since $v'$ needs the coefficient of message $j'$ to be 1 and $v$ requires the coefficient of message $j'$ to be 0. This last case contributes additional degree of at most $(d_1 - 1)d_2$. □

**Algorithm 3 for $t > 1$:** For the cases where the reduction from P3 to MWIS is polynomial, we can directly use an MWIS solver to solve the problem. Hence, we focus on the case that the reduction is not polynomial. We design an algorithm that applies for general $t$ and operates in $t$ iterations. In each iteration, the algorithm simply solves one instance of an MWIS problem that is reduced from a P3 sub-problem instance with 1 transmission. We present the algorithm in Alg. 3. In the first iteration, we solve the MWIS described earlier to select the transmission the server makes. Next, we update the problem instance: (i) we add decoded messages into side-information sets, and (ii) if client $i$ has decoded message $j$, we set $w_{ij'} = \max\{0, w_{ij'} - w_{ij}\}$ for all $j' \in R(i)$, to reflect the additional benefit that receiving message $j'$ would bring to client $i$ given that she has already received $j$. We proceed with the next iteration by solving the MWIS

---

**Algorithm 3** Greedy Coding Algorithm to Solve P3.

---

1: **Input**: number of messages $m$, number of clients $n$, request sets $R_i, \forall i \in [n]$, weights $w_{i,j}, \forall i \in [n], j \in R_i$, number of transmissions $t$.
2: **Output**: $t$ encoded transmissions $\{x_1, x_2, \ldots, x_t\}$.
3: **Initialization**: problem instance $\mathcal{I} = (m, n, \{R_i\}_{i \in [n]}, \{w_{ij}\}_{i \in [n], j \in R_i})$, benefit for client $i$ $s_i = 0, \forall i \in [n]$.
4: **for** $\tau = 1 : t$ **do**
5:     Map the problem instance $\mathcal{I}$ with one transmission into an MWIS instance $\mathcal{J}$.
6:     Solve the MWIS problem $\mathcal{J}$ and get output $\{j_1, j_2, \ldots, j_l\}$.
7:     Set the $\tau$-th encoded transmission to be $x_\tau = b_{j_1} + b_{j_2} + \ldots + b_{j_l}$.
8:     Update the instance $\mathcal{I}$:
9:     **if** Client $i$ ($\forall i \in [n]$) can decode message $j \in R_i$ **then**
10:         Move $j$ from $R_i$ to $S_i$.
11:         **if** $w_{ij} > 0$ **then**
12:             Update the individual benefit: $s_i = s_i + w_{ij}$ (received plus additional benefits).
13:             Set $w_{ij'} = 0$ for all $j' \in R_i$ and $w_{ij'} \leq w_{ij}$.
14:             Set $w_{ij'} = w_{ij'} - w_{ij}$ for all $j' \in R_i$ and $w_{ij'} > w_{ij}$.
15:         **end if**
16:     **end if**
17: **end for**

---

problem on the new instance. Observe that this scenario admits the index coding problem as a special case, and hence we can show that the P3 problem is hard to approximate within a ratio of $n^{1-\epsilon}$ for any $\epsilon > 0$ (see Appendix C).

## VI. NUMERICAL EVALUATION

### A. Over Random Instances

For Figs. 3-10 we uniformly at random generate instances with the parameters described in the captions, and present values averaged over all instances. In the following experiments, we normalize the benefit $B$ to get $B_0$ (divided by the maximum possible benefit to have maximum value 1).

**Trade-off between $B$ and $t$:** Figs. 3 and 4 show for P1 (Alg. 1) and P2 (Alg. 2) the trade-off between the number of transmissions $t$ and the normalized benefit $B_0$. We consistently observe that we can achieve a large percentage of the benefit with a small fraction of the transmissions we need to achieve the maximum benefit. For example, in Fig. 3, a 2% decrease in benefit can achieve a 71% bandwidth savings, and in Fig. 4, a 20% decrease in benefit can achieve a 91% bandwidth savings.

**Benefit for small $t$:** Figs. 5 and 6 highlight the normalized benefit $B_0$ we can achieve if the server is restricted to very few transmissions ($t = 1$ or $t = 4$) over some scenarios. Observe that with $t = 4$ transmissions we can consistently achieve more than 85% of the benefit and with $t = 1$ more than 38% of the benefit.

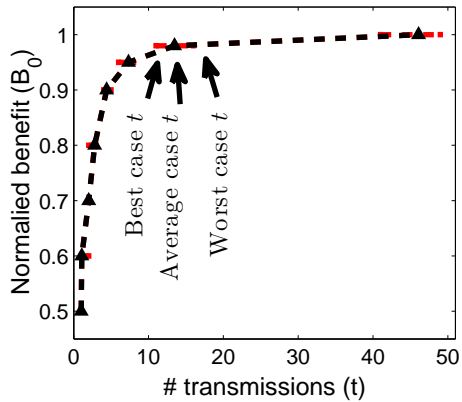

Fig. 3: Trade-off between bandwidth $t$ and normalized benefit $B_0$ for Alg. 1 and P1 with $m = 300$, $n = 50$ and Borda count model, averaged over 100 random instances.

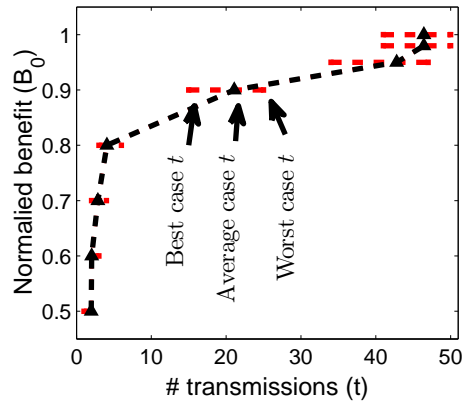

Fig. 4: Trade-off between bandwidth $t$ and normalized benefit $B_0$ for Alg. 2 and P2 with $m = 300$, $n = 50$, $k = 60$ and Borda count model, averaged over 100 random instances.

TABLE I: Description of scenarios

| Scenarios | Side information | Preference model | Parameters | Algorithms |
|---|---|---|---|---|
| Scenario 1 | No | Borda count | $m = 1000$, $n = 20$ | Alg. 1 |
| Scenario 2 | No | Bimodal benefit with $G = 10$ and $F = 0.1$ | $m = 1000$, $n = 20$ | Alg. 1 |
| Scenario 3 | Yes, $k = 100$ | Borda count | $m = 1000$, $n = 20$ | Alg. 2 |
| Scenario 4 | Yes, $k = 100$ | Bimodal benefit with $G = 10$ and $F = 0.1$ | $m = 1000$, $n = 20$ | Alg. 3 |

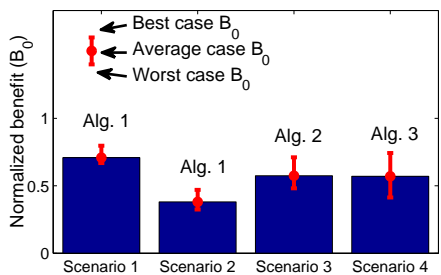

Fig. 5: Normalized benefit for $t = 1$. The scenarios are described in Table I.

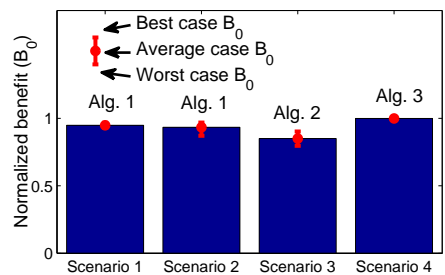

Fig. 6: Normalized benefit for $t = 4$. The scenarios are described in Table I.

**Benefit from coding:** Figs. 7 and 8 compare, over two sets of parameters for P2, the performance of Alg. 1 (we run Alg. 1 by ignoring the side information and making uncoded transmissions) and Alg. 2. We find that leveraging the side information and coding enables Alg. 2 to double in some cases the benefit.

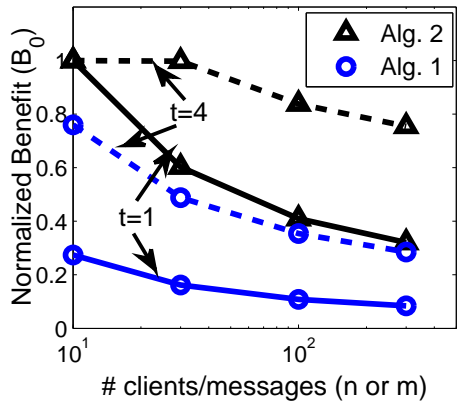

Fig. 7: We consider instances of P2 with $n = m$, $k = 0.1m$, and the Borda count model.

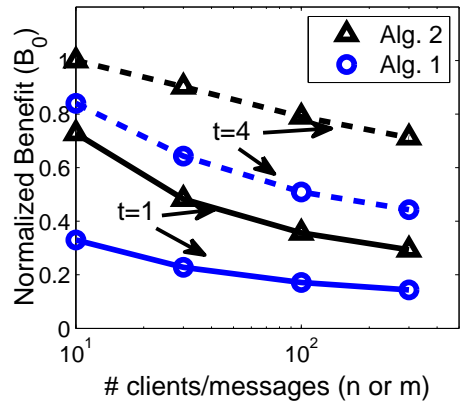

Fig. 8: We consider instances of P2 with $n = m$, $k = 0.2m$, and the Borda count model.

**Benefit over random selection:** Figs. 9 and 10 compare the performance of Alg. 1 with that of random selection over a modified instances of P1 with bimodal benefit model. Random selection assumes that the server first identifies all messages that form first preference for at least one client, and then randomly selects to transmit $t$ of them. We find that Alg. 1 achieves $52\% - 80\%$ more benefit than random selection for $t = 4$, and $60\% - 88\%$ more for $t = 1$ as $m$ changes from 10 to 1000; and achieves $30\% - 71\%$ more benefit than random selection for $t = 4$, and $31\% - 106\%$ more for $t = 1$ as $G$ changes from 2 to 50.

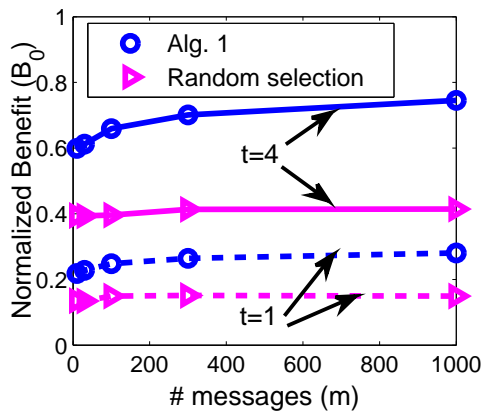

Fig. 9: We consider P1 with $n = 50$, and the bimodal benefit model with $G = 10$ and $F = 0.1$.

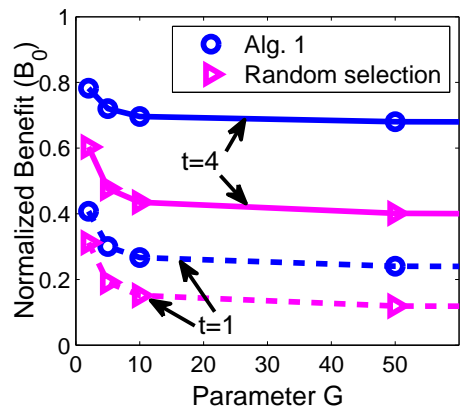

Fig. 10: We consider P1 with $m = 300, n = 50$, and the bimodal benefit model with $F = 0.1$, as $G$ varies.

*B. Over Data Sets*

We extract instances from the Yahoo! Search Marketing advertiser bidding data set, which was collected every $15$ minutes over a year's period; data instances include the time stamp, the key phrase ID, the advertiser ID, and the bidding price [11]. We generate a problem instance as follows: During each hour, $n$ users with $n$ search query key phrases enter the system (these are the clients). The advertiser bids to place an ad (the ads are the messages) to some of the key phrases using certain prices (we assume price zero for the rest of the key phrases). We interpret the price of a bid for a key phrase as the benefit of this ad (message) with respect to the key phrase (client). We assume that the messages that the advertiser does not bid for, form side information.

We compare Alg. 1 and 3 with the conventional Borda count method [10], the Spearfoot rule-based method [17], [19], and the Kemeny's method [18], [19] (we interpret the recommendations these algorithms make as the uncoded messages to transmit). The horizontal axis represents time (instances collected at sequential time slots). In Figs. 11 and 13 the vertical axis represents the actual benefit achieved at each time (current instance); in Figs. 12 and 14, the vertical axis represents the accumulated benefit (all previous instances). We find that all uncoded algorithms (including Alg. 1) perform similarly; this is because in the data set a few of the messages are the highest preferred by all clients, and thus the preference model used by the algorithm did not make a difference in the message choice. However, by leveraging the side information, Alg. 3 could accrue multiple times the benefit over time.

## VII. DISCUSSION, OPEN PROBLEMS AND CONCLUSIONS

In this paper, we have examined the recommendation benefit under bandwidth constraint in an index coding framework. We presented three problem examples to show that although the problems are in general NP hard, designing polynomial time approximation algorithms can still make a significant bandwidth savings by leveraging coding. We also conducted experiments over real data set to validate our arguments. We consistently found that, if we take into account bandwidth constraints, we get trends of diminishing returns: we can achieve the majority of the possible benefits with a fraction of the transmissions required to achieve the optimal benefits, if we carefully select what to transmit.

Our work in this paper focused on single-round protocols, where a server is faced with a set of clients that may already have some side information; we implicitly assume that if a client has a message as side information, the message has been displayed to her and thus cannot continue to generate benefits in future rounds. Thus the client preferences are restricted to the messages unknown to her, as these are the only messages that can bring additional benefit. However, it would be interesting to examine multi-round protocols, where perhaps at a first round clients have no side information, and as the rounds progress, clients utilize received messages as side information. We note that we can alter our single-round protocol to operate over multiple rounds by adjusting the message preferences from round to round; however the question on how to operate over multiple rounds, potentially with clients leaving and joining a broadcasting server, remains open.

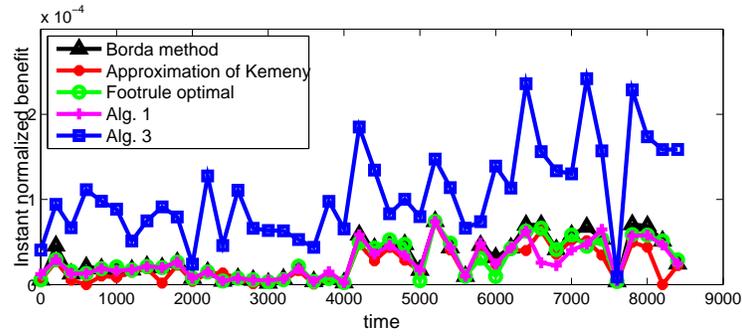

Fig. 11: Instant benefit for $t = 2$ as a function of time (current instance).

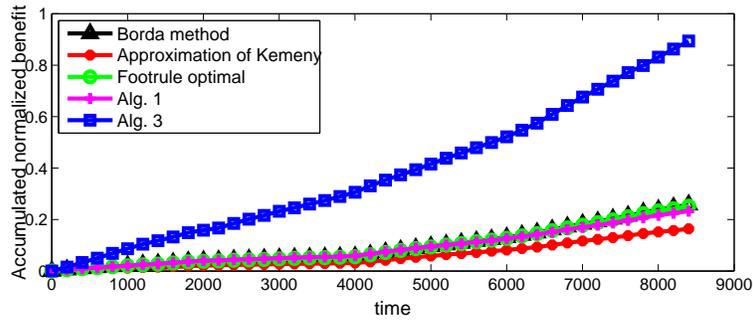

Fig. 12: Accumulated benefit for $t = 2$ over time (all previous instances).

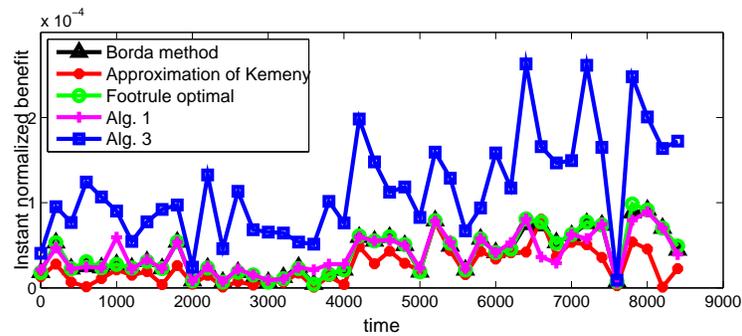

Fig. 13: Instant benefit for $t = 4$ as a function of time (current instance).

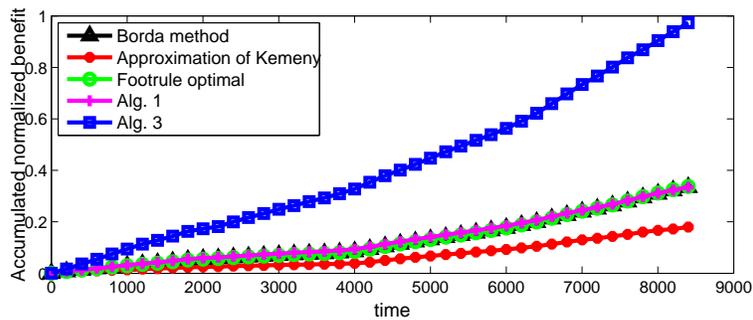

Fig. 14: Accumulated benefit for $t = 4$ as a function of time (all previous instances).

Our work selected some specific preference models and benefit aggregation rules to examine, but we believe that similar approaches could extend to alternative formulations as well, which are currently open questions. For example, an alternative rule to combine benefits is the sum of the benefits of all messages that clients receive, i.e., $s_i = \sum_{j \in D_i} s_i(j)$. We here discuss at a high level how in this case our algorithms could be adapted. Problem P1 becomes trivial under the sum benefit rule. It suffices to select the $t$ uncoded messages that achieve the maximum total benefit one by one, i.e., selecting the $t$ columns of the benefit matrix that have the $t$ largest sum of columns. The algorithm for solving P2 can also be modified for the sum benefits rule. The main modification would be, that in line 19 of Alg. 2, we discard a client during the transmission stages, if this client receives a sufficiently large benefit. However, when we consider the sum benefit rule, as long as the remaining unknown messages can give still benefit to this client, we do not need to discard her and can simply keep her. For P3, we can easily extend the heuristic Alg. 3 to create a heuristic algorithm that applies for the sum benefit case. Recall that we solve an MWIS problem instance at each iteration, and each solution corresponds to a transmission. For the P3 problem instance with the sum benefit rule, we can still construct the transmission scheme using an MWIS solver at each stage, but use a slightly different weight updating rule, that keep the benefits of each message the same independently of what other messages a client has received.

More generally, although in this paper we focused on some particular cases that we think are interesting, we believe that the general framework we have developed could be useful for other setups and problems as well, such as weighted pliable index coding, coded caching with preferences, and coded distributed computing.

APPENDIX A

APPENDIX FOR SECTION III (P1)

### A.1 *t-uncoded Transmissions Can Achieve Optimum*

We show that uncoded transmissions are sufficient to achieve the optimal benefit for problem P1.

In this paper, although we restrict our study to the case of linear schemes, the upper and lower bounds for P1 also apply even if we are allowed to use nonlinear schemes. This follows because (i) we have error free broadcast transmissions and (ii) clients have no side information, thus each client will receive exactly the same transmissions and be able to decode exactly the same messages. Thus intuitively, with $t$ transmissions, we cannot hope to decode more than $t$ messages.

More formally, let us consider the following general (linear or non-linear) encoding and decoding processes (for simplicity, we denote the output of the decoder as messages with indices $j_1, j_2, \ldots, j_{|D_i|}$).

$$\begin{aligned} \text{Encoding: } \phi_0(b_1, b_2, \ldots, b_m) &= (x_1, x_2, \ldots, x_t), \\ \text{Decoding: } \phi_i(x_1, x_2, \ldots, x_t) &= (b^\dagger_{j_1}, b^\dagger_{j_2}, \ldots, b^\dagger_{j_{|D_i|}}), \end{aligned} \tag{10}$$

where $b^\dagger_{j_1}, b^\dagger_{j_2}, \ldots, b^\dagger_{j_{|D_i|}}$ denotes the set of messages that client $i$ is able to decode. Note that in problem P1, $D_i$ is the same for all client $i \in [n]$.

Then we have the Markov chain $(b_1, b_2, \ldots, b_m) \rightarrow (x_1, x_2, \ldots, x_t) \rightarrow (b^\dagger_{j_1}, b^\dagger_{j_2}, \ldots, b^\dagger_{j_{|D_i|}})$. Using the data processing inequality, we have

$$\begin{aligned} I(b_1, b_2, \ldots, b_m; b^\dagger_{j_1}, b^\dagger_{j_2}, \ldots, b^\dagger_{j_{|D_i|}}) &\leq I(b_1, b_2, \ldots, b_m; x_1, x_2, \ldots, x_t) \\ &\leq H(x_1, x_2, \ldots, x_t) \leq t \log(q). \end{aligned} \tag{11}$$

This implies that the maximum number of messages that can be decoded is $|D_i| = t$. Moreover, if one client can decode a message, so do all other clients - there is nothing to differentiate among them. Thus, in this case it is optimal to use uncoded transmissions - there are no gains from linear or nonlinear coding schemes.

### A.2 *Proof of Theorem 1*

Consider a set cover problem instance, with a universe set $U = \{1, 2, \ldots, n\}$ and a family of subsets of $U$, $\mathcal{S} = \{S_1, S_2, \ldots, S_k\}$. The union of elements of $\mathcal{S}$ is $U$, i.e., $S_1 \cup S_2 \cup \ldots \cup S_k = U$ and $S_j \subseteq U$ for all $j \in [k]$. The goal of the set cover problem is to find a subset of $\mathcal{S}$, $\mathcal{S}' \subseteq \mathcal{S}$, with minimum cardinality such that $\cup_{S \in \mathcal{S}'} S = U$.

We show that a set cover problem instance can be reduced to a P1 problem instance with $m$ messages and $n$ clients in polynomial time. The clients will correspond to elements in the universe $U$, and we will have $m = (n+2)(nk-n+1)$ messages, that correspond to the $k$ subsets in the family $\mathcal{S}$ plus some additional (dummy) messages for construction purposes. We will construct a corresponding $n \times m$ benefit matrix $\Pi$, and sequentially test whether a selection of $t = 1, 2, \ldots, k$ columns can achieve a benefit greater than or equal to $n(m+1-k)$. Recall that in the benefit matrix the rows correspond to clients (elements of $U$), and the columns to messages (each one of the first $k$ columns will represent a subset in the family $\mathcal{S}$ and the remaining the dummy messages).

Let us denote by $\mathcal{S}[i]$ the family of subsets that contains element $i \in U$, i.e., $\mathcal{S}[i] = \{S \in \mathcal{S} : i \in S\}$. We denote the cardinality of $\mathcal{S}[i]$ by $d_i \leq k$ and the columns corresponding to subsets in $\mathcal{S}[i]$ by $j_{i1}, j_{i2}, \ldots, j_{id_i}$. Let us define $g = nk - n + 1$ and $d = d_1 + d_2 + \ldots + d_n$. Then $m = (n+2)g$. The benefit matrix we would like to construct has the following form:

$$\Pi = [m+1]_{n \times m} - \Pi^C = [m+1]_{n \times m} - [\Pi_0, \Pi_1, \Pi_2, \ldots, \Pi_n, R], \quad (12)$$

where $[m+1]_{n \times m}$ denotes the $n \times m$ matrix with all entries equal to $m+1$; $\Pi_0$ is of size $n \times k$; $\Pi_i$ $(1 \leq i \leq n)$ is of size $n \times (g - d_i)$; and $R$ is of size $n \times (2g - k + d)$. Note that for the matrix $\Pi^C$, each row is also a permutation of $[m]$ using the Borda count model. The construction process of $\Pi^C$ is as follows.

• Step 1: construct submatrix $\Pi_0$. We assign values row by row to the matrix. For the $i$-th row, assign an arbitrary permutation of values $1, 2, \ldots, d_i$ to $d_i$ positions corresponding to $\mathcal{S}[i]$, and assign an arbitrary permutation of values $g+1, g+2, \ldots, g+k-d_i$ to $(k-d_i)$ positions corresponding to $\mathcal{S}^C[i] = \mathcal{S} \backslash \mathcal{S}[i]$.

• Step 2: construct submatrix $\Pi_i$, $1 \leq i \leq n$, of size $n \times (g - d_i)$. First, assign an arbitrary permutation of the values $d_i+1, d_i+2, \ldots, g$ as the row vector of row $i$. For the other rows $l \neq i$, assign an arbitrary permutation of the values $(i+1)g+1, (i+1)g+2, \ldots, (i+1)g+(g-d_i)$ as the row vector of row $l$.

• Step 3: construct submatrix $R$. For each row $i$ of submatrix $R$, assign an arbitrary permutation of the remaining $(2g-k+d)$ values $g+k-d_i+1, g+k-d_i+2, \ldots, 2g, 2g+g-d_1+1, \ldots, 3g, 3g+$ $g - d_2 + 1, \ldots, 4g, \ldots, (n+1)g + g - d_n + 1, \ldots, (n+2)g, (i+1)g + 1, (i+1)g + 2, \ldots, (i+1)g + (g - d_i)$ as a row vector of row $i$.

*Example 1: We illustrate the construction of matrix $\Pi$ using a simple example. Let us consider a set cover problem represented by the following adjacency matrix:*

$$\begin{bmatrix} 1 & 1 & 0 & 1 \\ 0 & 1 & 1 & 0 \\ 1 & 0 & 1 & 0 \end{bmatrix},$$

*where each element of the universe $U = \{1, 2, 3\}$ is represented by a row and each subset in the subset family $\mathcal{S} = \{S_1 = \{1, 3\}, S_2 = \{1, 2\}, S_3 = \{2, 3\}, S_4 = \{1\}\}$ is represented by a column. In this case, we have $n = 3$, $k = 4$, $g = nk - n + 1 = 10$, $m = (n+2)g = 50$. Then we can construct our matrix $\Pi^C$ in the following form:*

$$\Pi^C = \begin{bmatrix} 1 & 2 & 11 & 3 & 4 & 5 & \ldots & 10 & 31 & 32 & \ldots & 38 & 41 & 42 & \ldots & 48 & 12 & \ldots \\ 11 & 1 & 2 & 12 & 21 & 22 & \ldots & 27 & 3 & 4 & \ldots & 10 & 41 & 42 & \ldots & 48 & 13 & \ldots \\ 1 & 11 & 2 & 12 & 21 & 22 & \ldots & 27 & 31 & 32 & \ldots & 38 & 3 & 4 & \ldots & 10 & 13 & \ldots \end{bmatrix}.$$

$$\qquad\qquad \Pi_0 \qquad\qquad\qquad \Pi_1 \qquad\qquad\qquad \Pi_2 \qquad\qquad\qquad \Pi_3 \qquad\qquad R$$

Next, we prove that finding a cover of size $t$ can be achieved by finding $t$ columns that achieve a certain benefit; hence we can sequentially test whether $t$ is the size of the minimum set cover, from which the theorem follows. We map a selection of $t$ sets for the set cover problem to the selection of the corresponding messages (columns of $\Pi_0$) to transmit; and reversely, when selecting messages to transmit, we consider subsets in the set cover problem corresponding to messages/columns of $\Pi_0$ (we ignore dummy messages). Thus we say that a selection of $t$ messages "covers" a client, if the corresponding set in the set cover problem includes this client.

**Lemma 2.** *The selection of $t$ subsets in $\mathcal{S}$ can cover the universe $U$, if and only if we can find $t$ columns in matrix $\Pi$ that achieve benefit at least $n(m + 1 - k)$.*

*Proof.* ● Necessity. If a selection of $t$ subsets in $\mathcal{S}$ can cover the universe $U$, this selection can achieve at least a benefit of $m + 1 - k$ for any client, resulting in a total benefit at least $n(m + 1 - k)$ according to the construction of $\Pi_0$ in step 1.

- Sufficiency. Suppose there exists an optimal selection $\mathcal{T}$ of size $t$ that does not cover some client $i \in [n]$ and achieves a benefit $B_\mathcal{T} \geq n(m + 1 - k)$.

We observe that the selection $\mathcal{T}$ does not contain any column in $\Pi_i$.

Indeed, if the selection $\mathcal{T}$ does not cover client $i$, but contains a column $j$ in $\Pi_i$, then we can simply find another selection strategy by replacing column $j$ with the column $j'$ that contains the value 1 for client $i$ in $\Pi_0$ (i.e., $s_i(j') = m + 1$). Formally, we can construct $\mathcal{T}' = \mathcal{T} \cup \{s_i^{-1}(m + 1)\} \backslash \{j\}$, where $s_i^{-1}(m + 1) = j'$ is the column that contains the most preferred message for client $i$ with benefit $m + 1$. From the construction, we can see that the benefits that message $j'$ can bring are larger than $j$ for all clients and at least client $i$ can improve her benefit. So $\mathcal{T}'$ is a better selection than $\mathcal{T}$, resulting in a contradiction with $\mathcal{T}$ being optimal.

Thus it suffices to consider the case that $\mathcal{T}$ does not cover client $i$ and does not contain any column in $\Pi_i$. From the construction, we note that columns corresponding to $\mathcal{S}[i]$ and $\Pi_i$ contain choices of values $1, 2, \ldots, g$ for client $i$. As a result, the maximum benefit client $i$ can achieve is $m - g$. Therefore, the benefit can be bounded by

$$B_\mathcal{T} \leq (n - 1)m + (m - g) = n(m + 1 - k) - 1 < n(m + 1 - k), \tag{13}$$

which results in a contradiction with $B_\mathcal{T} \geq n(m + 1 - k)$.

We finally argue that a selection $\mathcal{T}$ of $t$ columns of the benefit matrix $\Pi$ that can cover all $i \in U$, does not contain columns outside $\Pi_0$. This follows from the construction, since a column $j_1$ in $\Pi_i$ $(i = 1, 2, \ldots, n)$ is dominated by any column in $\Pi_0$ that can cover client $i$ (i.e., $j_1$ gives a lower benefit for any client compared with any column in $\Pi_0$), such that the removal of $j_1$ will not affect the benefit. This means that $\mathcal{T} \backslash \{j_1\}$ can achieve the same benefit as $\mathcal{T}$, if $\mathcal{T}$ contains at least a selection in $\Pi_0$. In this case, the testing would stop at most at step $t - 1$. Similarly, any column $j_2$ in $R$ is dominated by any column in $\Pi_0$ such that the removal of $j_2$ will not affect the benefit.  □

*Example 2: Continuing with our previous example 1, if we choose $t$ columns, we can see that only when $t$ columns covering at least one column in $\{1, 2, 4\}$ (i.e., covering element 1), at least one column in $\{2, 3\}$ (i.e., covering element 2), and at least one column in $\{1, 3\}$ (i.e., covering element 3) we can make the benefit no less than $n(m + 1 - k) = 141$).*

*A.3 Proof of Theorem 2*

We use $B(\tau)$ and $\delta(\tau)$, $\tau = 1, 2, \ldots, t$, to denote the benefit collected after $\tau$ steps and the increase of benefit in the $\tau$-th step, respectively, using Alg. 1. That is:

$$
\begin{aligned}
B(1) &= \delta(1); \\
B(2) &= B(1) + \delta(2); \\
&\vdots \\
B_{\mathcal{T}} &= B(t) = B(t-1) + \delta(t).
\end{aligned}
\tag{14}
$$

We set $B(0) = 0$ and denote by $B^*$ the optimal benefit and by $\mathcal{T}^*$ an optimal selection of $t$ messages (columns of the benefit matrix) to transmit that achieve $B^*$. For all $\tau$, we can bound $\delta(\tau)$ using the following lemma.

**Lemma 3.** *In Alg. 1, we can bound the increase in benefit in each step by:*

$$
\delta(\tau) \geq \frac{B^* - B(\tau - 1)}{t}, \quad \tau = 1, 2, \ldots, t.
\tag{15}
$$

*Proof.* To prove this lemma, we first observe the following. Consider two selections of messages (columns of the ranking matrix) $\mathcal{T}_1$ and $\mathcal{T}_2$, with $\mathcal{T}_1 \subseteq \mathcal{T}_2$. If we add a new column $j$ to these selections, we get that:

$$
B_{\mathcal{T}_1 \cup \{j\}} - B_{\mathcal{T}_1} \geq B_{\mathcal{T}_2 \cup \{j\}} - B_{\mathcal{T}_2}.
\tag{16}
$$

Indeed, when we add message $j$ (column $j$) to $\mathcal{T}_1$, assume $C$ is the set of clients whose benefit can be improved. The left hand side of eq. (16), denoted by $\delta_1$, is the additional benefit message $j$ can bring to the client in $C$ and to a level determined by their ranking of message $j$. In contrast, when we add message $j$ (column $j$) to $\mathcal{T}_2$, clients not in $C$ cannot receive additional benefit, because we have $\mathcal{T}_1 \subseteq \mathcal{T}_2$. Clients in $C$, however, can receive additional benefit at most $\delta_1$, since some of the clients in $C$ may have already received higher benefit compared with the selection $\mathcal{T}_1$ due to that $\mathcal{T}_1 \subseteq \mathcal{T}_2$).

Going back to the proof of (15), observe that at the beginning of the $\tau$-th step, the difference from the optimal benefit is $B^* - B(\tau - 1)$. From the pigeonhole principle, at least one of the $t$ messages in the optimal set $\mathcal{T}^*$, let us say $j$, can improve the benefit at least by $\frac{B^* - B(\tau-1)}{t}$, since a future selection of $j$ can only offer improvements less than or equal to what it can offer in the current stage from (16). $\qquad\square$

Using Lemma 3 we can prove the bound of our theorem.

$$
\begin{aligned}
B(t) \ &= B(t-1) + \delta(t) \\
&\geq B(t-1) + \tfrac{B^* - B(t-1)}{t} \\
&= (1 - \tfrac{1}{t})B(t-1) + \tfrac{1}{t}B^* \\
&\geq (1 - \tfrac{1}{t})[(1 - \tfrac{1}{t})B(t-1) + \tfrac{1}{t}B^*] + \tfrac{1}{t}B^* \\
&= (1 - \tfrac{1}{t})^2 B(t-2) + \tfrac{1}{t}[1 + (1 - \tfrac{1}{t})]B^* \\
&\geq \ldots \\
&\geq (1 - \tfrac{1}{t})^{t-1}\tfrac{B^*}{t} + \tfrac{1}{t}[1 + (1 - \tfrac{1}{t}) + (1 - \tfrac{1}{t})^2 + \ldots + (1 - \tfrac{1}{t})^{t-2}]B^* \\
&= (1 - \tfrac{1}{t})^{t-1}\tfrac{B^*}{t} + [1 - (1 - \tfrac{1}{t})^{t-1}]B^* \\
&= [1 - (1 - \tfrac{1}{t})^t]B^* \\
&\geq [1 - \tfrac{1}{e}]B^*,
\end{aligned}
\tag{17}
$$

where the first four inequalities hold by repeatedly using Lemma 3; the second to the last equality holds according to the finite geometric series calculation; the last inequality holds according to the inequality $(1 - \tfrac{1}{t})^t \leq 1/e$.

### A.4  Proof of Theorem 3

**Lower bound:** We prove the lower bound by showing that a random selection of $t$ columns achieves the expected benefit $\mu \triangleq \frac{tn(m+1)}{t+1}$ and thus, there exists a selection of $t$ columns that achieves at least such benefit, $B \geq \mu$.

Consider an $n \times m$ benefit matrix $\Pi$. Assume we select uniformly at random $t$ columns from all $m$ columns, i.e., with equal probability select 1 from all $\binom{m}{t}$ possible choices. We will calculate the expected benefit from this random selection.

For a given selection, if we denote by $X_1, X_2, \ldots, X_n$ the benefit received by clients $1, 2, \ldots, n$, then the total benefit is $B = X_1 + X_2 + \ldots + X_n$. Note that since we randomly select the $t$ columns and the rows can have arbitrary assignments of (Borda count) benefits for each client, the expected benefits are equal for each client, namely, $\mathbb{E}X_1 = \mathbb{E}X_2 = \ldots = \mathbb{E}X_n$. Thus it is sufficient to just calculate $\mathbb{E}X_1$.

Denote by $\mathcal{T}$ a set of $t$ indices for columns. We next consider all $(m - t + 1)$ possibilities for the benefit $X_1$ that client 1 has received. We will use the notation $s_i^{-1}(x)$ to refer to the message that brings benefit $x$ to client $i$.

• Case 1: the most preferred message is in $\mathcal{T}$, i.e., $s_1^{-1}(m) \in \mathcal{T}$, and in this case we have $X_1 = m$. The probability of $\{s_1^{-1}(m) \in \mathcal{T}\}$ is $p_1 = \frac{t}{m}$.

• Case 2: the message with benefit $m$ is in $[m]\backslash\mathcal{T}$ and the message with benefit $m-1$ is in $\mathcal{T}$, i.e., $\{s_1^{-1}(m-1) \in \mathcal{T} \wedge s_1^{-1}(m) \notin \mathcal{T}\}$. In this case, $X_1 = m-1$ and occurs with probability $p_2 = \frac{t(m-t)}{m(m-1)}$, where $\frac{t}{m}$ is the probability of $\{s_1^{-1}(m-1) \in \mathcal{T}\}$ and $\frac{m-t}{m-1}$ is the probability of $\{s_1^{-1}(m) \notin \mathcal{T}\}$.

Continuing along these lines we get:

• Case $j$: the messages with benefits $m, m-1, \ldots, m-j+2$ are in $[m]\backslash\mathcal{T}$ and the message with benefit $m+1-j$ is in $\mathcal{T}$, i.e., $\{s_1^{-1}(m+1-j) \in \mathcal{T} \wedge s_1^{-1}(m) \notin \mathcal{T} \wedge \ldots \wedge s_1^{-1}(m-j+2) \notin \mathcal{T}\}$. In this case, $X_1 = m+1-j$, and occurs with probability $p_j = \frac{t(m-t)(m-t-1)\ldots(m-t-j+2)}{m(m-1)(m-2)\ldots(m-j+1)}$, where $\frac{t}{m}$ is the probability of $\{s_1^{-1}(m+1-j) \in \mathcal{T}\}$, and $\frac{(m-t)(m-t-1)\ldots(m-t-j+2)}{(m-1)(m-2)\ldots(m-j+1)}$ is the probability of $\{s_1^{-1}(m) \notin \mathcal{T} \wedge \ldots \wedge s_1^{-1}(m-j+2) \notin \mathcal{T}\}$. Hence, we have

$$
\begin{aligned}
\mathbb{E}X_1 &= p_1 m + p_2(m-1) + \ldots + p_j(m+1-j) + \ldots + p_{m-t+1}t \\
&= \frac{t}{m!}\sum_{j=1}^{m-t+1}[P(m-t,j-1)(m-j+1)!] = \frac{t(m+1)}{t+1},
\end{aligned}
\tag{18}
$$

where the third equality holds from Lemma 4 that we provide later in this appendix.

**Upper bound:** We use a probabilistic method to construct a $n \times m$ benefit matrix $\Pi$ as follows. Draw a permutation of $[m]$ from all $m!$ possible permutations i.i.d. uniformly at random, and assign it to the $i$-th row of the ranking matrix $\Pi$, for all $i = 1, 2, \ldots, n$.

Consider a fixed selection of $t$ columns, e.g., $\mathcal{T} = [t] = \{1, 2, \ldots, t\}$. Let $X_1, X_2, \ldots, X_n$ be the benefits received by clients $1, 2, \ldots, n$, then the total benefit is $B = X_1 + X_2 + \ldots + X_n$. Due to the i.i.d. uniform selection for each row of $\Pi$, we have that $\mathbb{E}X_1 = \mathbb{E}X_2 = \ldots = \mathbb{E}X_n$. Therefore, We only need to calculate $\mathbb{E}X_1$, by listing all the possibilities for client 1's (first row of the benefit matrix) received benefit. Recall that we use $s_1^{-1}(x)$ to denote the column (message) that can bring benefit $x$ to client 1. We have the following $(m-t+1)$ possibilities for $X_1$.

• Case 1: the message with benefit $m$ is in $[t]$, i.e., $\{s_1^{-1}(m) \in [t]\}$. In this case $X_1 = m$, and the probability of this event is $p_1 = \frac{t(m-1)!}{m!}$, where $t$ is the number of ways to assign benefit $m$ to a message in $[t]$; $(m-1)!$ is the number of ways to assign benefits for the remaining $(m-1)$ messages; and the total number of possible assignments is $m!$. We underline that each benefit assignment (a permutation of $[m]$) for the row occurs with equal probability.

• Case 2: the message with benefit $m$ is in $[m]\backslash[t]$ and the message with benefit $m-1$ is in $[t]$, i.e., $\{s_1^{-1}(m-1) \in [t] \wedge s_1^{-1}(m) \notin [t]\}$. In this case, $X_1 = m-1$, and the associated probability is $p_2 = \frac{tP(m-t,1)(m-2)!}{m!}$, where $t$ is the number of ways to assign benefit $m-1$ to a message in $[t]$; $P(m-t,1)$, the 1-permutation of $m-t$, is the number of ways to assign benefit $m$ to a message in the remaining $m-t$ positions $t+1, t+2, \ldots, m$; $(m-2)!$ is the number of ways to assign benefits for the remaining $(m-2)$ messages; and the total number of possible assignments is $m!$.

Continuing along these lines we get:

• Case $j$: the messages with benefits $m, m-1, \ldots, m-j+2$ are in $[m]\backslash[t]$ and the message with benefit $m+1-j$ is in $[t]$, i.e., $\{s_1^{-1}(m+1-j) \in [t] \wedge s_1^{-1}(m) \notin [t] \wedge \ldots \wedge s_1^{-1}(m-j+2) \notin [t]\}$. In this case, $X_1 = m+1-j$, and this event occurs with probability $p_j = \frac{tP(m-t,j-1)(m-j)!}{m!}$, where $t$ is the number of ways to assign benefit $m+1-j$ to a message in $[t]$; $P(m-t,j-1)$, the $j$-permutations of $m-t$, is the number of ways to assign benefits $m, m-1, \ldots, m-j+2$ to $j-1$ messages in the remaining $m-t$ positions $t+1, t+2, \ldots, m$; $(m-j)!$ is the number of ways to assign benefits for the remaining $(m-j)$ messages; and the total number of possible assignments is $m!$. Therefore:

$$
\begin{aligned}
\mathbb{E}X_1 &= p_1 m + p_2(m-1) + \ldots + p_j(m+1-j) + \ldots + p_{m-t+1}t \\
&= \frac{t}{m!} \sum_{j=1}^{m-t+1} [P(m-t, j-1)(m-j+1)!] = \frac{t(m+1)}{t+1},
\end{aligned} \tag{19}
$$

where the third equality holds from Lemma 4.

From the Chernoff bound, we can bound the probability that the benefit is above $B_{UPPER} = (1+\delta)\mu$ (where $\delta = \sqrt{\frac{6t\log(m)}{n}}$):

$$
\Pr\{B \ge (1+\delta)\mu\} = \Pr\{\tfrac{B}{m} \ge \tfrac{1}{m}(1+\delta)\mu\} \le e^{-\frac{\mu\delta^2}{3m}} = e^{-\frac{2t^2(m+1)\log(m)}{m(t+1)}} \triangleq \epsilon, \tag{20}
$$

where the first equality is to normalize the random variable $B$, i.e., $\frac{B}{m} = \frac{X_1}{m} + \frac{X_2}{m} + \ldots + \frac{X_n}{m}$, such that $\frac{X_i}{m}$ is between $0$ and $1$. Note that $\mathbb{E}\frac{B}{m} = \frac{\mu}{m}$.

The inequality $\Pr\{B \ge (1+\delta)\mu\} \le \epsilon$ implies that for the fixed selection of $t$ columns, $[t]$, there are at most an $\epsilon$ fraction of instances when selecting the matrix $\Pi$ that can achieve a benefit no less than $(1+\delta)\mu$. Due to the uniformly at random selection of $\Pi$, given any fixed selection of $t$ columns, it is also the case that at most $\epsilon$ fraction of instances can achieve this benefit $(1+\delta)\mu$. There are in total $\binom{m}{t}$ possible selections of columns. The fraction of instances of matrices $\Pi$ that can achieve a benefit no less than $(1+\delta)\mu$ given any of the $\binom{m}{t}$ selections is at most

$$\binom{m}{t}\epsilon < m^t e^{-\frac{2t^2(m+1)\log(m)}{m(t+1)}} = e^{(t\log(m)-\frac{2t^2(m+1)\log(m)}{m(t+1)})} = e^{(-\frac{t\log(m)[(t-1)m+2t]}{m(t+1)})} < 1, \qquad (21)$$

which indicates that there must exist instances, such that, for any selection of $t$ columns, the average benefit cannot be more than $B_{UPPER}$. This concludes the proof of the theorem.

**Lemma 4.**

$$\frac{1}{m!}\sum_{j=1}^{m-t+1}[P(m-t,j-1)(m-j+1)!] = \frac{m+1}{t+1}, \textit{for any } t \le m. \qquad (22)$$

*Proof.* Change the variable of this equation by setting $k = m - t$. Denote by $H(k)$ the expression on the left hand side of eq. (22), i.e.,

$$H(k) = \frac{1}{m!}\sum_{j=1}^{k+1}[P(k,j-1)(m-j+1)!]. \qquad (23)$$

To show that $H(k) = \frac{m+1}{m-k+1}$ for any $m \ge k$, we use a mathematical induction method for $k$ and consider $m$ as a parameter.

For $k = 0$, $\frac{1}{m!}P(0,0)m! = 1$ and for $k = 1$, $\frac{1}{m!}(P(1,0)m! + P(1,1)(m-1)!) = 1 + \frac{1}{m}$, eq. (22) holds. Assume eq. (22) holds for $k \le m - 1$. Now, for $k+1$, we have

$$\begin{aligned}
H(k+1) &= \frac{1}{m!}\sum_{j=1}^{k+2}[P(k+1,j-1)(m-j+1)!] \\
&= \frac{1}{m!}[\frac{(k+1)!}{(k+1)!}m! + \sum_{j=2}^{k+2}\frac{(k+1)!}{(k+1-j+1)!}(m-j+1)!] \\
&= 1 + \frac{k+1}{m(m-1)!}\sum_{l=1}^{k+1}\frac{k!}{(k+1-l)!}(m-1-l+1)! \\
&= 1 + \frac{(k+1)}{m}\cdot\frac{m-1+1}{m-1-k+1} \\
&= \frac{m+1}{m-k},
\end{aligned} \qquad (24)$$

where the third equality holds due to a change of variable $l = j - 1$ and the fourth equality holds due to the induction hypothesis on $k$ with parameter $m - 1$. Therefore, the equation (22) holds. □

*A.5 Proof of Theorem 4*

We consider a family of instances $\mathcal{I}$, each with $m$ messages and $n$ clients. The benefit matrix $\Pi$ is generated as follows: for each row, uniformly and independently draw a permutation from all $m!$ permutations of $[m]$ and assign it to the row.

We first show that $\Delta(2, n, 1) = \frac{1}{\sqrt{2\pi n}}$ for even $n$ and $\Delta(2, n, 1) = \frac{1}{\sqrt{2\pi(n-1)}}$ for odd $n \geq 3$.

Consider a $n \times 2$ benefit matrix instance $\Pi$ in the family $\mathcal{I}$. Define $n_1$ and $n_2$ to be the numbers of 1s and 2s in the first column. Obviously, we have $n_1 + n_2 = n$ and $n_1$ and $n_2$ are also the numbers of 2s and 1s in the second column. Our strategy is to select column 2 if $n_1 \geq n_2$ and to select column 1, otherwise. For even $n$, the expected benefit with respect to $\Pi$ is

$$\mathbb{E}B = \sum_{n_1=0}^{n/2} \frac{1}{2^n} \binom{n}{n_1} [2(n - n_1) + n_1] + \sum_{n_1=n/2+1}^{n} \frac{1}{2^n} \binom{n}{n - n_1} [(n - n_1) + 2n_1], \quad (25)$$

where the first term corresponds to the selection of column 1 and the second term corresponds to the selection of column 2. Here, in the first term, $\binom{n}{n_1}$ is the number of choices for $n_1$ 2s in the second column, resulting in a probability of $\frac{1}{2^n} \binom{n}{n_1}$ that the benefit is $2(n - n_1) + n_1$. The interpretation is similar for the second term. Similarly, for odd $n$, the expectation of benefit with respect to $\Pi$ is

$$\mathbb{E}B = \sum_{n_1=0}^{\frac{n-1}{2}} \frac{1}{2^n} \binom{n}{n_1} [2(n - n_1) + n_1] + \sum_{n_1=\frac{n+1}{2}}^{n} \frac{1}{2^n} \binom{n}{n - n_1} [(n - n_1) + 2n_1]. \quad (26)$$

By simplifying this expression (see Lemma 5), we get that:

$$\mathbb{E}B = \begin{cases} \frac{3n}{2} + \frac{n}{\sqrt{2\pi n}}, & n \text{ even}, \\ \frac{3n}{2} + \frac{n}{\sqrt{2\pi(n-1)}}, & n \geq 3, \text{ odd}. \end{cases} \quad (27)$$

We next consider the general term $\Delta(m, n, t)$. The strategy we use here is as follows. We randomly select $t$ columns. If this selection can achieve a benefit no less than $\mu = \frac{tn(m+1)}{t+1}$, we keep these columns as our selection. If not, we discard these columns, and select columns with a benefit at least $\mu$. This is always possible according to Theorem 3.

Next, we look at when the benefit we actually achieve is greater than $\mu + \frac{n}{\sqrt{2\pi n}} \sigma(m, t)$.

For a fixed selection of $t$ columns, if $X_i$ is the benefit for client $i$, then the total benefit $X$ can be represented as $X = \sum_{i=1}^{n} X_i$. According to the central limit theorem, the distribution of $Y = \frac{X - \mu}{\sqrt{n}\sigma(m,t)}$ is approximately the standard normal distribution $\mathcal{N}(0, 1)$. Given our selection algorithm, the benefit is lower bounded by:

$$B \geq \begin{cases} \mu, & X \leq \mu, \\ X, & X > \mu. \end{cases} \tag{28}$$

Hence, the expected benefit can be lower bounded by

$$\mathbb{E}B \geq \mu + \Pr\{X > \mu\}\mathbb{E}[X - \mu | X > \mu]. \tag{29}$$

The second term $\Pr\{X > \mu\}\mathbb{E}[X - \mu | X > \mu]$ can be approximately calculated as:

$$\Pr\{X > \mu\}\mathbb{E}[X - \mu | X > \mu] \approx \int_0^\infty \sqrt{n}\sigma(m,t)y\phi(y)dy = \frac{n}{\sqrt{2\pi n}}\sigma(m,t), \tag{30}$$

where $\phi(y) = \frac{1}{\sqrt{2\pi}}e^{-\frac{y^2}{2}}$ is the probability density function of the standard normal distribution. The calculation of $\sigma(m,t)$ is in Lemma 6, from which the theorem follows.

In addition, we can strictly lower bound the average performance by using the Berry-Esseen theorem.

**Corollary 1.** *The expectation of benefit can be strictly lower bounded by* $\mu + \frac{n}{\sqrt{2\pi n}}\sigma(m,t)$ $(1 - e^{-\frac{m^2}{2}}) - \frac{m^3}{2\sigma^2(m,t)}$.

*Proof.* The Berry-Esseen theorem states that the difference of distribution of $Y = \frac{(X-\mu)}{\sqrt{n}\sigma(m,t)}$ and the normal distribution can be bounded by

$$|F_Y(y) - \Phi(y)| \leq \frac{\rho}{2\sigma^3(m,t)\sqrt{n}}, \tag{31}$$

for all $n$ and $y$, where $F_Y(y)$ and $\Phi(y)$ are the cumulative distribution functions of $Y$ and the normal distribution, and $\rho = \mathbb{E}|X/n - \mathbb{E}X/n|^3 < m^3$. We also know that $|X - \mu| \leq mn$. Hence, the term $\Pr\{X > \mu\}\mathbb{E}[X - \mu | X > \mu]$ can be bounded by

$$\begin{aligned} \Pr\{X > \mu\}\mathbb{E}[X - \mu | X > \mu] \quad &\geq \int_0^m \sqrt{n}\sigma(m,t)y\phi(y)dy - \sqrt{n}\sigma(m,t)\frac{\rho}{2\sigma^3(m,t)\sqrt{n}} \\ &= \frac{n}{\sqrt{2\pi n}}\sigma(m,t)(1 - e^{-\frac{m^2}{2}}) - \frac{m^3}{2\sigma^2(m,t)}. \end{aligned} \tag{32}$$

This proves the corollary. $\qquad\qquad\qquad\qquad\qquad\qquad\qquad\qquad\qquad\qquad\qquad \square$

**Lemma 5.**

$$\mathbb{E}B = \begin{cases} \frac{3n}{2} + \frac{n}{\sqrt{2\pi n}}, & n \text{ even}, \\[2mm] \frac{3n}{2} + \frac{n}{\sqrt{2\pi(n-1)}}, & n \geq 3, \text{ odd}. \end{cases} \tag{33}$$

*Proof.* When $n$ is even, we have

$$
\begin{aligned}
\mathbb{E}[B] &= \frac{1}{2^n} \Big[ \sum_{n_1=0}^{n/2} \binom{n}{n_1}(2(n-n_1)+n_1) + \sum_{n_2=0}^{n/2} \binom{n}{n_2}(2(n-n_2)+n_2) - \binom{n}{n/2}(\frac{3n}{2}) \Big] \\
&= \frac{2}{2^n} \Big[ \sum_{n_1=0}^{n/2} \binom{n}{n_1}(2n-n_1) - \binom{n}{n/2}(\frac{3n}{4}) \Big] \\
&= \frac{2}{2^n} \Big[ n(2^n + \binom{n}{n/2}) - n2^{(n-2)} - \binom{n}{n/2}(\frac{3n}{4}) \Big] \\
&\approx \frac{3n}{2} + \frac{n}{\sqrt{2\pi n}},
\end{aligned}
\tag{34}
$$

where the last approximation is due to the Stirling's approximation and the third equality holds because of the following two equations:

$$2^n = \sum_{n_1=0}^{n/2} \binom{n}{n_1} + \sum_{n_1=n/2+1}^{n} \binom{n}{n_1} = 2\sum_{n_1=0}^{n/2} \binom{n}{n_1} - \binom{n}{n/2}, \tag{35}$$

and

$$\sum_{n_1=0}^{n/2} \binom{n}{n_1} n_1 = n\sum_{n_1=1}^{n/2} \frac{(n-1)!}{(n_1-1)!(n-n_1)!} = n\sum_{k=0}^{n/2-1} \frac{(n-1)!}{k!(n-1-k)!} = n2^{(n-2)}. \tag{36}$$

When $n \geq 3$ is odd, we have

$$
\begin{aligned}
\mathbb{E}[B] &= \frac{1}{2^{(n-1)}} \Big[ \sum_{n_1=0}^{(n-1)/2} \binom{n}{n_1}(2n-n_1) \Big] \\
&= \frac{1}{2^{(n-1)}} \Big[ n2^n - (n2^{(n-2)} - \frac{n}{2}\binom{n-1}{(n-1)/2}) \Big] \\
&\approx \frac{3n}{2} + \frac{n}{\sqrt{2\pi(n-1)}},
\end{aligned}
\tag{37}
$$

where the last approximation is due to the Stirling's approximation and the second equality holds because of the following two equations:

$$2^n = \sum_{n_1=0}^{(n-1)/2} \binom{n}{n_1} + \sum_{n_1=(n+1)/2}^{n} \binom{n}{n_1} = 2\sum_{n_1=0}^{(n-1)/2} \binom{n}{n_1}, \tag{38}$$

and

$$\sum_{n_1=0}^{(n-1)/2} \binom{n}{n_1} n_1 = n\sum_{n_1=1}^{(n-1)/2} \frac{(n-1)!}{(n_1-1)!(n-n_1)!} = n\sum_{k=0}^{(n-3)/2} \binom{n-1}{k} = n\frac{2^{(n-1)} - \binom{n-1}{(n-1)/2}}{2}. \tag{39}$$

$\square$

**Lemma 6.**

$$\sigma(m,t) = \sqrt{\frac{(m+1)(m-t)t}{(t+1)^2(t+2)}} \tag{40}$$

*Proof.* We already know that the expected value of the benefit $X_1$ is $\frac{t(m+1)}{t+1}$, and the distribution of $X_1$ is as follows:

$$\Pr\{X_1 = m+1-j\} = \frac{tP(m-t, j-1)(m-j)!}{m!}, \; j = 1, 2, \ldots, m+1-t. \tag{41}$$

Next, we prove the following result using induction:

$$\mathbb{E}[X_1^2] \; = \sum_{j=1}^{m+1-t} \Pr\{X = m+1-j\}(m+1-j)^2 = \frac{t^2(m+1)}{t+1} + \frac{t(m-t)(m+1)}{t+2}, \tag{42}$$

or

$$\frac{1}{m!}\sum_{j=1}^{k+1}[P(k, j-1)(m-j+1)!(m-j+1)] = \frac{t(m+1)}{t+1} + \frac{(m-t)(m+1)}{t+2}. \tag{43}$$

We change the variable of this equation by setting $k = m-t$. We denote by $L(k)$ the following expression:

$$L(k) = \frac{1}{m!}\sum_{j=1}^{k+1}[P(k, j-1)(m-j+1)!(m-j+1)]. \tag{44}$$

When $k = 0$, the initial condition holds, i.e., $L(0) = m = \frac{m(m+1)}{m+1} + \frac{0(m+1)}{m-k+2}$. Assume that $L(k) = \frac{(m-k)(m+1)}{m-k+1} + \frac{k(m+1)}{t+2}$ holds for all $m > k$. Then, for $k+1$, we have

$$
\begin{aligned}
L(k+1) \; &= \frac{1}{m!}\sum_{j=1}^{k+2}[P(k+1, j-1)(m-j+1)!(m-j+1)] \\
&= \frac{1}{m!}m!m + \frac{1}{m!}\sum_{j=2}^{k+2}[\frac{(k+1)!}{(k+1-j+1)!}(m-j+1)!(m-j+1)] \\
&= m + \frac{k+1}{m}[\frac{1}{(m-1)!}\sum_{l=1}^{k+1}[\frac{k!}{(k+1-l)!}(m-1-l+1)!(m-1-l+1)]] \\
&= m + \frac{k+1}{m}[\frac{(m-1-k)m}{m-k} + \frac{km}{m-k+1}] \\
&= \frac{(m-k-1)(m+1)}{m-k} + \frac{(k+1)(m+1)}{m-k+1},
\end{aligned}
\tag{45}
$$

where the third equality holds due to a change of variable $l = j - 1$ and the fourth equality holds due to the induction hypothesis on $k$ with parameter $m - 1$. Therefore eq. (42) is proved.

Furthermore, we can calculate $\sigma^2(m,t) = \mathbb{E}[X_1^2] - \mu^2 = \frac{(m+1)(m-t)t}{(t+1)^2(t+2)}$. $\qquad\square$

APPENDIX B

APPENDIX FOR SECTION IV (P2)

**Derandomization Function for Claim 1:** We here describe a polynomial-time deterministic algorithm to select a coding vector $\boldsymbol{a}^\xi$. We refer to the clients that can decode a message that gives benefit at least $k + 1 - \xi$ (they have ranked less than or equal to $\xi$) as the *qualified clients*. For a given row coding vector $\boldsymbol{a}$, we denote the number of qualified clients by $Y[\boldsymbol{a}]$.

We sequentially assign a coding coefficient $0$ or $1$ to the $m$ coding coefficients in the vector $\boldsymbol{a} = (a_1, a_2, \ldots, a_m)$ in $m$ steps. At the beginning of the $j$-th step, the first $j - 1$ coefficients have been assigned some values $a_1 = \bar{a}_1, \ a_2 = \bar{a}_2, \ldots, a_{j-1} = \bar{a}_{j-1}$.

We define $Y_{\bar{\boldsymbol{a}}_{[j-1]}, 0} = \mathbb{E}_{\boldsymbol{a}} Y[\boldsymbol{a} | a_1 = \bar{a}_1, a_2 = \bar{a}_2, \ldots, a_{j-1} = \bar{a}_{j-1}, a_j = 0]$ to be the expected number of qualified clients, averaged over all coding vectors with the assigned values for the first $j - 1$ coding coefficients $\bar{\boldsymbol{a}}_{[j-1]}$ and a $0$ for the $j$-th coding coefficient; $Y_{\bar{\boldsymbol{a}}_{[j-1]}, 1} = \mathbb{E}_{\boldsymbol{a}} Y[\boldsymbol{a} | a_1 = \bar{a}_1, a_2 = \bar{a}_2, \ldots, a_{j-1} = \bar{a}_{j-1}, a_j = 1]$ to be the expected number of qualified clients, averaged over all coding vectors with the assigned values for the first $j - 1$ coding coefficients $\bar{\boldsymbol{a}}_{[j-1]}$ and a $1$ for the $j$-th coding coefficient; and $Y_{\bar{\boldsymbol{a}}_{[j]}} = \mathbb{E}_{\boldsymbol{a}} Y[\boldsymbol{a} | a_1 = \bar{a}_1, a_2 = \bar{a}_2, \ldots, a_j = \bar{a}_j]$ to be the expected number of qualified clients, averaged over all coding vectors with the assigned values $\bar{\boldsymbol{a}}_{[j]}$ for the first $j$ coefficients. We also denote by $\boldsymbol{a}_{[0]}$ the empty set.

The algorithm proceeds as follows: *For step $j = 1, 2, \ldots, m$, assign the $j$-th coding coefficient $\bar{a}_j$ to be $1$ if $Y_{\bar{\boldsymbol{a}}_{[j-1]}, 1} \geq Y_{\bar{\boldsymbol{a}}_{[j-1]}, 0}$ and $0$ otherwise.*

In the derandomization function in Alg. 4, the steps 10-14 essentially implement the calculation of the expected values we use for the decision making. We track the probability that a client $i$ will be a qualified client, $p_i^{j-1}$, for each step $j - 1$. Then we choose the coding coefficient $a_j$ by comparing the expected numbers of qualified clients if we choose a $0$ and a $1$ for $a_j$. We set a state parameter $z_i^j$ to represent the number of coefficient assignment choices for the remaining messages in $R_i$ such that client $i$ can remain qualified. For example, $z_i^j = 3$ if $|\{j' \in R_i | j' > j, \pi_i(j) \leq \xi, a_{j''} = 0, \forall j'' \in R_i \text{ and } j'' \leq j\}| = 3$.

**Derandomization function performance:** We here argue that the coding vector $\bar{\boldsymbol{a}}$ we identify enables at least $\frac{n\xi}{ke}$ clients to be qualified, i.e., decode a message in their request set that gives benefit at least $k + 1 - \xi$. Let $Y_{\bar{\boldsymbol{a}}}$ be the number of qualified clients after running the derandomized function, and let $Y$ be the average number of qualified clients after the randomized selection in Claim 1, where we i.i.d. at random assigned value $0$ to each coding coefficient with probability $p$. By tracking the number of qualified clients, originally we have $Y = (1 - p) Y_{\emptyset, 0} +$ $pY_{\emptyset,1}$; then $Y_{\emptyset,0} \geq Y$ or $Y_{\emptyset,1} \geq Y$ holds; and hence we have $Y_{\hat{a}_{[1]}} \geq Y$. For step $j$, we can see that $Y_{\hat{a}_{[j]}} = (1-p)Y_{\hat{a}_{[j-1]},0} + pY_{\hat{a}_{[j-1]},1}$. Hence, at least one of the two following inequalities, $Y_{\hat{a}_{[j-1]},0} \geq Y_{\hat{a}_{[j]}}$ and $Y_{\hat{a}_{[j-1]},0} \geq Y_{\hat{a}_{[j]}}$, holds. Therefore, using the derandomization function, we have $Y_{\hat{a}_{[j]}} \geq Y_{\hat{a}_{[j-1]}}$. Hence, $Y_{\hat{a}} \geq Y \geq \frac{n\xi}{ke}$ holds.

## Appendix C

## Appendix for Section V (P3)

We here discuss the hardness of approximation for P3. Note that P3 admits the index coding as a special case, where each client requires one message with benefit 1 and others with benefit 0. The hardness of approximating index coding capacity is shown through a reduction from the MIS problem [21]. Here, we use a similar reduction to show the hardness of approximating $B^*$ as follows.

**Proposition 1.** *The P3 problem is hard (unless $NP = ZPP$) to approximate within a ratio of $n^{1-\epsilon}$ for any $\epsilon > 0$.*

*Proof.* For this we use the result that the MIS is hard (unless $NP = ZPP$) to approximate within a ratio of $n^{1-\epsilon}$ for any $\epsilon > 0$ [22]. We map an MIS instance on a graph $\mathcal{G} = (V, E)$, into a P3 problem as follows (for completeness).

• We create a P3 instance with $m = n = |V|$: we map each of the $|V|$ vertices in $\mathcal{G}$ to a client in P3; and we also create $|V|$ messages.

• A client $i$ has message $i$ in her request set with individual benefit $w_{ii} = 1$.

• A client $i$ has a message $j \neq i$ in her request set $R_i$ with individual benefit $w_{ij} = 0$ if and only if there is an edge between vertex $i$ and vertex $j$ in $\mathcal{G}$. All the remaining messages are in the side information set of client $i$.

We next argue that the size of the maximum independent set in $\mathcal{G}$ equals the maximum benefit that we can achieve over the constructed P3 instance if we are restricted to $t = 1$ transmission. Indeed, given an independent set $\mathcal{S}$ in $\mathcal{G}$, we can construct a row coding vector for P3 that enables each client $i$ in $\mathcal{S}$ to decode message $i$ (we simply use coding coefficients 1 for all such $i \in \mathcal{S}$ and 0 for the remaining coding coefficients). Recall that with one transmission, a client can decode a message $i$ in her request set if and only if the coding coefficient for this message is nonzero and the coding coefficients for all other messages in her request set are zero. This would achieve benefit $|\mathcal{S}|$ in P3.

**Algorithm 4** Derandomization Function.

1: **Input**: number of messages $m$, number of clients $n$, request sets $R_i, \forall i \in [n]$, individual benefit $s_i(j), \forall i \in [n], j \in R_i$, size of request set $k$, and threshold $\xi$.

2: **Output**: row coding vector $\boldsymbol{a} \in \{0,1\}^m$.

3: **Initialization**: set the client set $\mathcal{N} = [n]$; set the qualification probability $p_i^0 = \frac{\xi}{k}(1-\frac{1}{k})^{k-1}$, for all client $i \in [n]$; set the state $z_i^0 = \xi$ for each client $i \in [n]$.

4: **for** $j = 1 : m$ **do**

5:     **for** all $i \in \mathcal{N}$ **do**

6:         **if** $j \notin R_i$ **then**

7:             $p_{i,0}^j = p_{i,1}^j = p_i^{j-1}$;   $z_{i,0}^j = z_{i,1}^j = z_i^{j-1}$. //A client is not affected if not connected to message $j$.

8:         **end if**

9:         **if** $j \in R_i$ and $s_i(j) \geq k+1-\xi$ **then**

10:             // This is the case that client $i$ receives at least $k+1-\xi$ benefit from message $j$.

              // Update the probability that client $i$ can be qualified if $a_j$ is 0 or 1:

$$p_{i,0}^j = \begin{cases} 0, & \text{if } z_i^{j-1}=1 \text{ and } a_{j'}=0 \text{ for all } j'<j \text{ and } j' \in R_i, \\ \frac{p_i^{j-1}}{1-1/k}, & \text{if } z_i^{j-1}=1 \text{ and } a_{j'}=1 \text{ for some } j'<j \text{ and } j' \in R_i, \\ \frac{(1-1/z_i^{j-1})p_i^{j-1}}{1-1/k}, & \text{otherwise;} \end{cases} \quad (46)$$

$$p_{i,1}^j = \begin{cases} 0, & \text{if } a_{j'}=1 \text{ for some } j'<j \text{ and } j' \in R_i, \\ \frac{kp_i^{j-1}}{z_i^{j-1}}, & \text{otherwise.} \end{cases} \quad (47)$$

11:             // Update the state of client $i$ if $a_j$ is 0 or 1:

$$z_{i,0}^j = \begin{cases} 0, & \text{if } z_i^{j-1}=1 \text{ and } a_{j'}=0 \text{ for all } j'<j \text{ and } j' \in R_i, \\ 1, & \text{if } z_i^{j-1}=1 \text{ and } a_{j'}=1 \text{ for some } j'<j \text{ and } j' \in R_i, \\ z_i^{j-1}-1, & \text{otherwise;} \end{cases} \quad (48)$$

$$z_{i,1}^j = \begin{cases} 0, & \text{if } a_{j'}=1 \text{ for some } j'<j \text{ and } j' \in R_i, \\ 1, & \text{otherwise.} \end{cases} \quad (49)$$

12:         **end if**

13:         **if** $j \in R_i$ and $s_i(j) < k+1-\xi$ **then**

14:             // This is the case that client $i$ receives less than $k+1-\xi$ benefit from message $j$.

              // Update the probability that client $i$ can be qualified if $a_j$ is 0 or 1:

$$p_{i,0}^j = \frac{p_i^{j-1}}{1-1/k}; \quad p_{i,1}^j = 0. \quad (50)$$

15:             // Update the state of client $i$ if $a_j$ is 0 or 1:

$$z_{i,0}^j = z_i^j; \quad z_{i,1}^j = 0. \quad (51)$$

16:         **end if**

17:     **end for**

18:     **if** $\sum_{i \in \mathcal{N}: j \in R_i} p_{i,1}^j \geq \sum_{i \in \mathcal{N}: j \in R_i} p_{i,0}^j$ **then**

19:         Set $a_j = 1$, $p_i^j = p_{i,1}^j$, and $z_i^j = z_{i,1}^j$.

20:     **else**

21:         Set $a_j = 0$, $p_i^j = p_{i,0}^j$, and $z_i^j = z_{i,0}^j$.

22:     **end if**

23:     Remove $i$ from $\mathcal{N}$, if $z_i^j = 0$ for all $i \in \mathcal{N}$.

24: **end for**

We argue that this is the maximum benefit we could achieve in $P3$: indeed, if a larger benefit was possible in $P3$, more than $|\mathcal{S}|$ clients would have been able to decode their corresponding requested messages with benefit 1. Also note that, a row coding vector that achieves a maximum benefit $B^*$, enables $B^*$ clients to decode their corresponding requested messages with benefit 1, and thus directly determines an independent set of size $|\mathcal{S}| = B^*$ in $\mathcal{G}$. $\qquad\square$